\documentclass[12pt]{iopart}

\usepackage{hyperref}
\usepackage{cleveref}
\usepackage{graphicx}
\usepackage{amssymb}
\usepackage[english]{babel}
\usepackage{soul}
\usepackage{color}
\usepackage[normalem]{ulem}
\usepackage[numbers]{natbib}

\usepackage{aas_macros}
\newcommand{\cqg}{Class.~Quant.~Grav.}

\begin{document}

\title{From the microscopic to the macroscopic world: from nucleons to neutron stars}

\author{Gandolfi, S.}
\address{Theoretical Division, Los Alamos National Laboratory, Los Alamos, New Mexico 87545, USA}
\author{Lippuner, J.}
\address{Computational Physics and Methods, CCS-2, Los Alamos National Laboratory, Los Alamos, New Mexico 87545, USA}
\address{Center for Theoretical Astrophysics, Los Alamos National Laboratory, Los Alamos, NM 87545, USA}
\author{Steiner, A. W.}
\address{Department of Physics and Astronomy, University of
  Tennessee, Knoxville, TN 37996, USA}
\address{Physics Division, Oak Ridge National Laboratory, Oak
  Ridge, TN 37831, USA}
\author{Tews, I.}
\address{Theoretical Division, Los Alamos National Laboratory, Los Alamos, New Mexico 87545, USA}
\author{Du, X.}
\address{Department of Physics and Astronomy, University of
  Tennessee, Knoxville, TN 37996, USA}
\author{Al-Mamun, M.}
\address{Department of Physics and Astronomy, University of
  Tennessee, Knoxville, TN 37996, USA}

\begin{abstract}
Recent observations of neutron-star properties, in particular the recent detection of gravitational waves emitted from binary neutron stars, GW 170817,
open the way to put strong constraints on nuclear interactions.
In this paper, we review the state of the art in calculating the
equation of state of strongly interacting matter from first principle
calculations starting from microscopic interactions among nucleons.
We then review selected properties of neutron stars that can be
directly compared with present and future observations.
\end{abstract}

\section{Introduction:}

The idea that astronomical observations may provide insight into
nuclear interactions originated in the
1950s~\cite{Burbidge57,Cameron59}. There have been a few landmark
neutron-star observations in the past decade which can be directly
connected to the interaction between nucleons. Two neutron-star mass
measurements obtain results $\sim 2$ solar
masses~\cite{Demorest:2010,Antoniadis:2013} (see also
\cite{Fonseca16}). Simultaneous information on neutron-star masses and
radii is becoming available~\cite{Ozel10am,Steiner10te} (see updates
in Refs.~\cite{Ozel16td,Steiner18ct,Nattila17ns}). Finally, the
first detection of gravitational waves from a binary neutron-star merger, GW 170817, has also provided mass
and tidal deformability constraints~\cite{ligo:17nsns}, and its electromagnetic counterpart has
demonstrated that neutron-star mergers are an important source for
r-process nuclei. These advances in neutron-star
observations provide a unique opportunity to improve our knowledge of
nuclear physics. Although the corresponding length scales are
separated by many orders of magnitude, properties of neutron stars and
nuclei are strongly connected. In particular, the equation of state
(EOS) of the crust and the outer core is one of the main ingredients
for neutron-star structure, determining radii, tidal deformabilities,
and other properties of neutron stars. The EOS is obviously related to
nuclear forces and properties of nuclei.

\begin{figure}[h!]
\begin{center}
\includegraphics[width=0.99\textwidth]{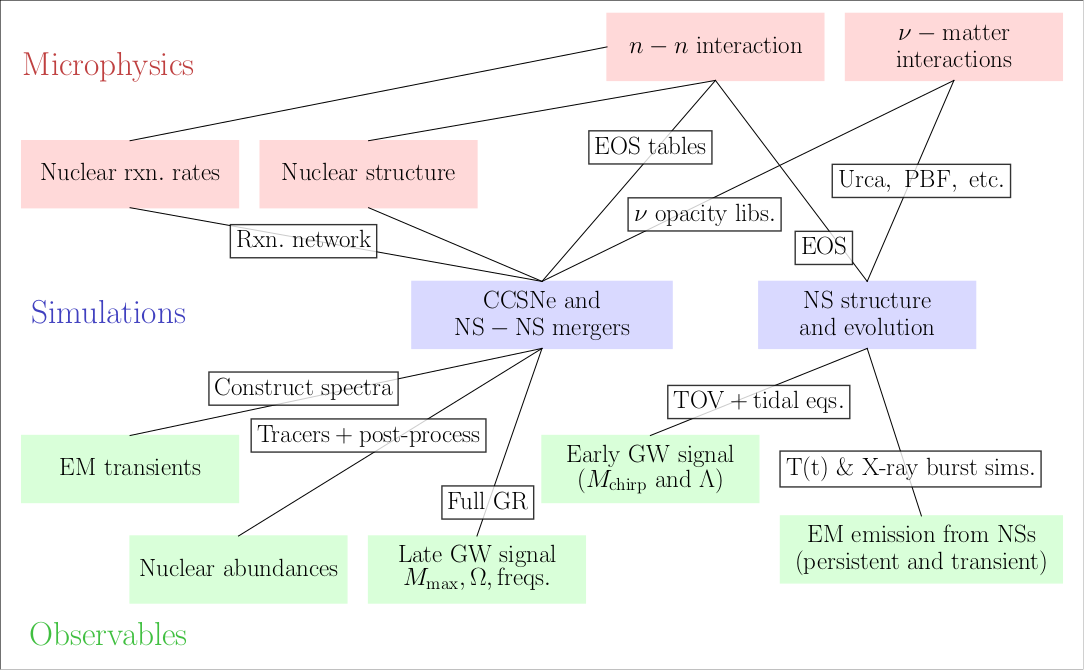}
\caption{A diagram representing the connections between
microphysical models and astrophysical observations.}
\label{fig:overview}
\end{center}
\end{figure}

At the femtometer scale, it is of fundamental importance to understand
how nucleons interact, how neutrinos interact in nuclear matter, how
nuclei are formed and how their properties emerge. All these are among the main
ingredients for astrophysical simulations, including simulations of supernova
explosions and the evolution of neutron stars. Those simulations
predict or explain observations of phenomena which can have exameter
(100 light-year) length scales, including electromagnetic and
gravitational-wave emission, nuclear abundances, ejecta from binary
neutron-star mergers, and others; see Fig.~\ref{fig:overview} for a
diagram summarizing the connections between microphysical models,
simulations, and observations. There is a strong connection between
the microscopic and macroscopic world, separated by many orders of
magnitude. Information in Fig.~\ref{fig:overview} flows both ways:
microphysical models can be used to describe observations, and
observations constrain our knowledge of nuclear interactions, nuclear
structure, and nuclear reactions.

In this paper, we will start by briefly discussing how nuclear forces
are constructed and tested, how they are used to calculate the EOS,
and then how the EOS is used to predict selected properties of neutron
stars. We will then provide details regarding how Bayesian inference
can be utilized to attack the ``inverse problem'': the problem of
constraining our model parameters from observational data.

\section{The EOS}\label{sec:EOS}

\begin{figure}
\begin{center}
\includegraphics[width=0.8\textwidth]{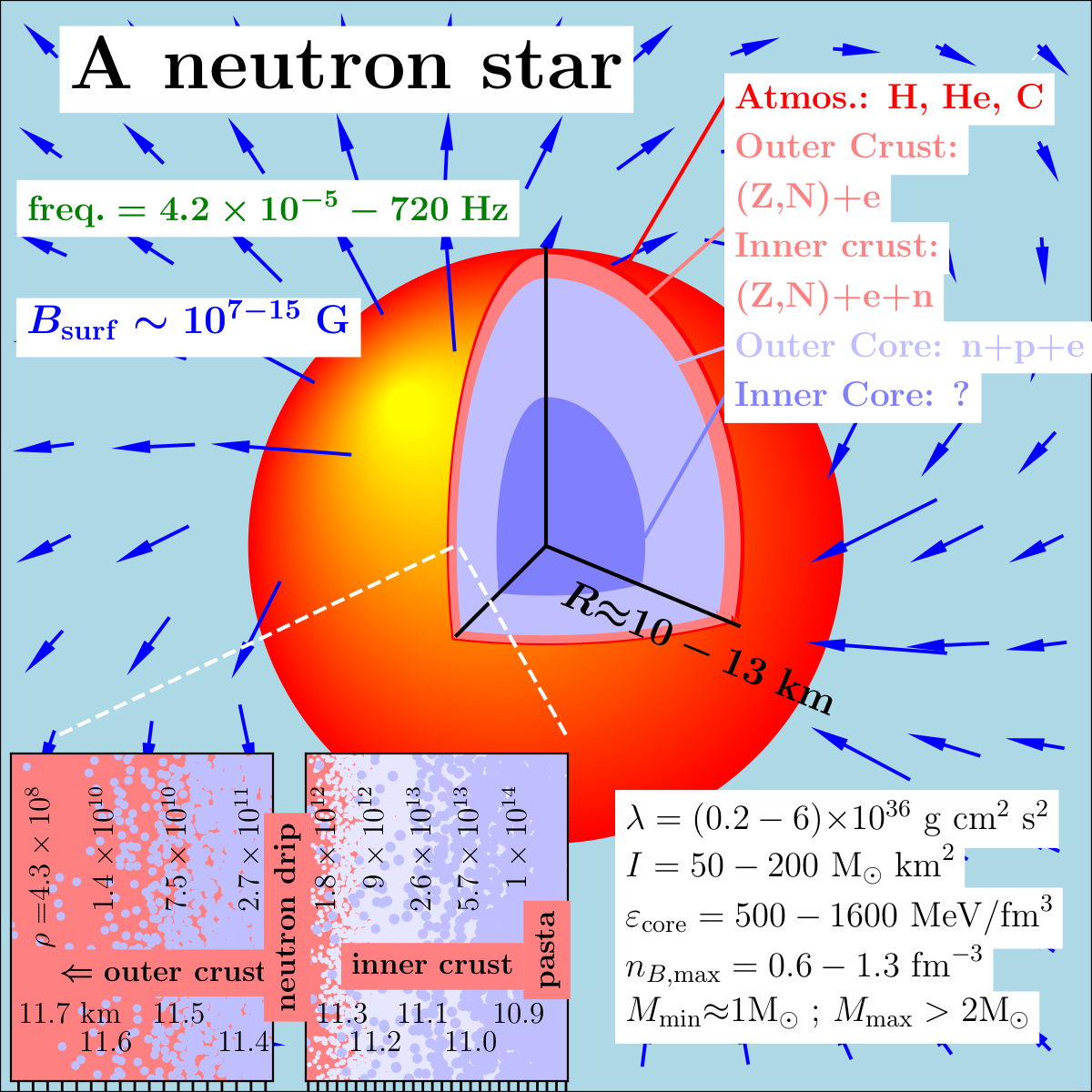}
\caption{A summary of the microphysics of neutron stars.
  In the upper left of the figure, the observational limits
  on rotation frequency~\cite{DAi16,Hessels06} and the
  magnetic field~\cite{Woods07,Karako-Argaman15} are given.
  The upper right panel shows the composition of the various layers.
  The lower left shows a schematic representation of the crust, where
  the dark blue color represents nuclei and the light blue color
  represents free neutrons. The limits on radius from X-ray
  observations are shown near the center~\cite{Steiner16ns}.
  Limits on the tidal deformability, moment of inertia,
  energy density in the core, and baryon density in the
  core~\cite{Steiner15un} are shown in the lower right panel.
  This figure
    was inspired by a previous version by Dany
    Page available at
    \href{http://www.astroscu.unam.mx/neutrones/NS-Picture/NStar/NStar.html}{http://www.astroscu.unam.mx/neutrones/NS-Picture/NStar/NStar.html}.
}
\label{fig:NS}
\end{center}
\end{figure}

Most of the static and dynamical properties of neutron stars can be
calculated once an Equation of State (EOS) describing the matter
inside the stars is specified. At very low densities, in the outer
crust of neutron stars, the matter is mainly composed of a lattice of
ordinary nuclei in the iron region. With increasing density, the
neutron chemical potential also increases and nuclei become extremely
neutron-rich. At the interface with the inner crust, the neutron
chemical potential is sufficiently high so that neutrons start to drip
out of the nuclei, and the nuclei start to be surrounded by a sea of
neutrons. Eventually, at the bottom of the inner crust, the geometry
of nuclei begins to be deformed, forming the so-called ``pasta''
phase. At about half the saturation density,
$\rho_0=0.16$~fm$^{-3}\approx2.7\cdot 10^{14}$~g/cm$^3$, the nuclei
completely melt, and the neutron-star core begins. Here, matter is
composed of a uniform liquid of a large fraction of neutrons with a
few protons, electrons, and eventually muons. At even higher
densities, above $\approx2\rho_0$, the composition of matter is
basically unknown. In this inner core of the neutron star, many
scenarios for the state of matter have been suggested, for example the
formation of hyperons~\cite{Lonardoni:2015}, quark
matter~\cite{Alford:2008}, or other more exotic condensates.
Fig.~\ref{fig:NS} shows the nature of these layers and gives a summary
of properties of a typical neutron star.

\begin{figure}
\begin{center}
\includegraphics[width=0.8\textwidth]{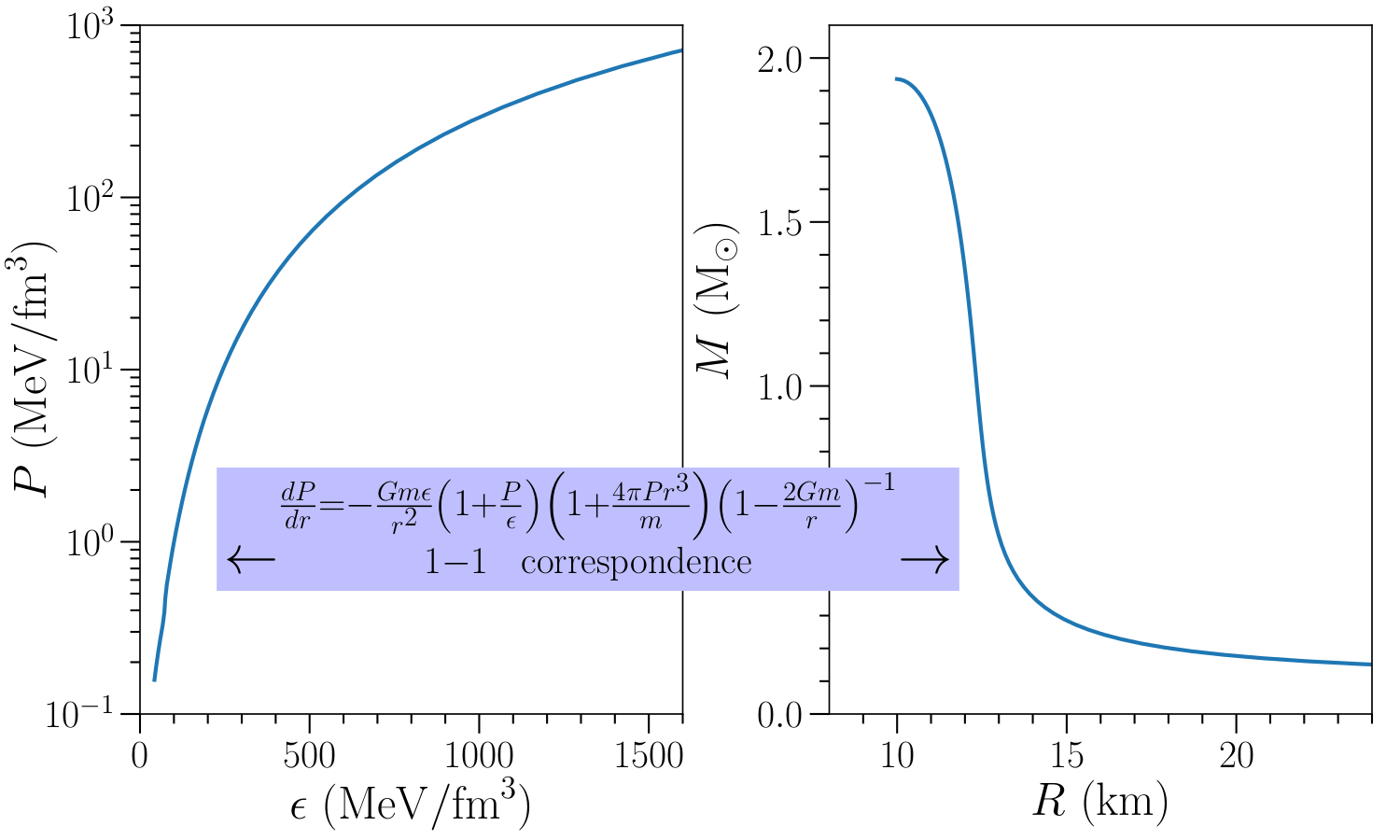}
\caption{The EOS and MR curve of Skyrme model NRAPR~\cite{Steiner05ia}.}
\label{fig:eos_mvsr}
\end{center}
\end{figure}

The full description of the EOS in the whole range of densities
encountered inside a neutron star, i.e.\ up to several times nuclear
saturation density, is a formidable task. Especially for the inner
core, most EOS used in astrophysical simulations necessarily make
some model-dependent assumptions. However, once an EOS $\epsilon(p)$
is specified, the mass-radius relation of a non-rotating neutron star
can be easily calculated by solving the Tolman-Oppenheimer-Volkoff
(TOV) equations:
\begin{eqnarray}
\frac{dP}{dr}&=-\frac{G[m(r)+4\pi r^3P/c^2][\epsilon+P/c^2]}{r[r-2Gm(r)/c^2]}\,,
\nonumber \\
\frac{dm(r)}{dr}&=4\pi\epsilon r^2 \,,
\end{eqnarray}
where $P$ and $\epsilon$
are the pressure and the energy density (including the
rest mass energy density contribution),
$m(r)$ is the gravitational mass enclosed within a radius $r$, and $G$
is the gravitational constant. The solution of the TOV equations for a
given central density gives the profiles of $\rho(r)$, $\epsilon(r)$
and $P(r)$ as functions of radius $r$, and also the total radius $R$
and mass $M=m(R)$, specified by the condition $P(R)=0$. By varying the
input central density, the mass-radius (MR) relation is mapped out. The
resulting MR relation is in one-to-one correspondence with the input
EOS (as summarized in Fig.~\ref{fig:eos_mvsr}),
which offers the possibility to measure the EOS, and thus the properties
of strongly interacting matter up to a few times saturation density, by
mapping out the MR relation observationally. In addition to properties of
isolated neutron stars, also other dynamical properties of neutron stars and
binary neutron-star mergers
depend on the EOS, including the gravitational-wave spectrum, the amount of ejected
material, and others.

An ideal starting point for the EOS of neutron-star matter is the EOS of
pure neutron matter (PNM), which is a homogeneous system that contains
only neutrons. The EOS of PNM can be calculated using sophisticated
many-body methods once a nuclear Hamiltonian is specified. Among
other methods, these include variational methods based on the cluster
expansion~\cite{Akmal:1998},
many-body perturbation theory~\cite{Hebeler:2010},
the coupled-cluster method~\cite{Hagen:2014}, and Quantum Monte Carlo
methods~\cite{Sarsa:2003,Carlson:2003}.
In this paper, we will focus on recent results obtained with
the Auxiliary Field Diffusion Monte Carlo (AFDMC) method, which was
originally introduced by Schmidt and Fantoni~\cite{Schmidt:1999}, and is ideally suited to study neutron matter~\cite{Gandolfi:2015,Carlson:2015}.

The main idea of QMC methods is to evolve
a many-body wave function in imaginary-time:
\begin{equation}
\Psi(\tau)=\exp\big[-H\tau\big]\,\Psi_v \,,
\end{equation}
where $\Psi_v$ is a variational ansatz of the many-body wave function
and $H$ is the Hamiltonian describing
the system. In the limit of $\tau\rightarrow\infty$, $\Psi$
approaches the ground-state of $H$. The evolution in imaginary-time
is performed by sampling configurations of the system using Monte
Carlo techniques, and expectation values are evaluated over
the sampled configurations. For more details see for example
Refs.~\cite{Carlson:2015,Pudliner:1997,Gandolfi:2009,Lonardoni:2018prl,
Lonardoni:2018prc,Lonardoni:2018nofk}.

In addition to the many-body method, one needs the nuclear Hamiltonian
as input, which describes the interactions among nucleons.  In AFDMC,
nuclei and neutron matter are described by non-relativistic point-like
particles interacting via two- and three-body forces:
\begin{equation}
H_{\rm nuc} = \sum_i\frac{p_i^2}{2m_N} + \sum_{i<j}v_{ij} + \sum_{i<j<k}v_{ijk} \,.
\end{equation}
The first two-body potential that has been extensively used with the AFDMC method is the phenomenological Argonne
AV8' potential~\cite{Wiringa:2002}, that is a simplified form of the Argonne
AV18 potential~\cite{Wiringa:1995}. Although simpler to use in QMC calculations,
the AV8' potential provides almost the same accuracy as AV18 in fitting NN
scattering data. In addition to such two-body potentials, it was shown that one has to include also a three-body interaction to being able to describe nuclear systems accurately, e.g., to correctly describe the binding energy
of light nuclei~\cite{Carlson:2015}. However, the three-body force is not as well constrained
as the NN interaction. The Urbana IX (UIX) three-body force has been originally proposed to be used in combination with the Argonne AV18 and AV8'~\cite{Pudliner:1995} potentials. Although
it slightly underbinds the energy of light nuclei, it has been
extensively used to study the equation of state of nuclear and neutron
matter~\cite{Akmal:1998,Gandolfi:2009,Gandolfi:2012}.
The AV8'+UIX Hamiltonian has been used in many works, and provided a
fairly good description of neutron star properties~\cite{Gandolfi:2012}.

However, such phenomenological interactions suffer from certain
shortcomings. Most importantly, they do not enable to estimate reliable
theoretical uncertainties and cannot be systematically improved. These
shortcomings can be addressed with the advent of chiral effective field
theory (EFT), which offers a systematic expansion of nuclear forces
that allows for theoretical uncertainties~\cite{Epelbaum:2009,Epelbaum:2015}.

In this approach, the relevant degrees of freedom are nucleons that can interact via explicit pion exchanges or via short-range contact interactions. The relevant diagrams entering in the nucleon-nucleon interaction are
systematically organized in powers of $p/\Lambda_b$, where $p$ is the
typical momenta of nucleons in the given nuclear system, i.e.\ similar
to the pion mass $m_\pi\approx 140$ MeV, and $\Lambda_b\approx 500-600$
MeV~\cite{Melendez:2017phj} is the so-called breakdown scale, where the chiral EFT expansion
is expected to loose its validity due to the increasing importance
of shorter-range physics, i.e., new degrees of freedom. Chiral interactions include long-range pion-exchange physics explicitly, while short-range physics is described by a
general operator basis consistent with all symmetries of the fundamental
theory, Quantum Chromodynamics.  The long-range pion-exchange terms
entering into the chiral EFT potentials are fully determined by
$\pi$-nucleon scattering data, while the parameters associated to
the contact terms, called low-energy constants (LECs), are typically
constrained by fitting nucleon-nucleon scattering data.  For more
details see Ref.~\cite{Epelbaum:2009}. Hamiltonians from
chiral EFT have been recently combined with QMC
methods~\cite{Gezerlis:2013,Gezerlis:2014,Lynn:2016,Lonardoni:2018,Lonardoni:2018prl}, and give very reasonable predictions of properties of
nuclei up to A=16, including energies,
radii, and momentum distributions, and neutron-$\alpha$ scattering.

\begin{figure}
\begin{center}
\includegraphics[width=0.65\textwidth]{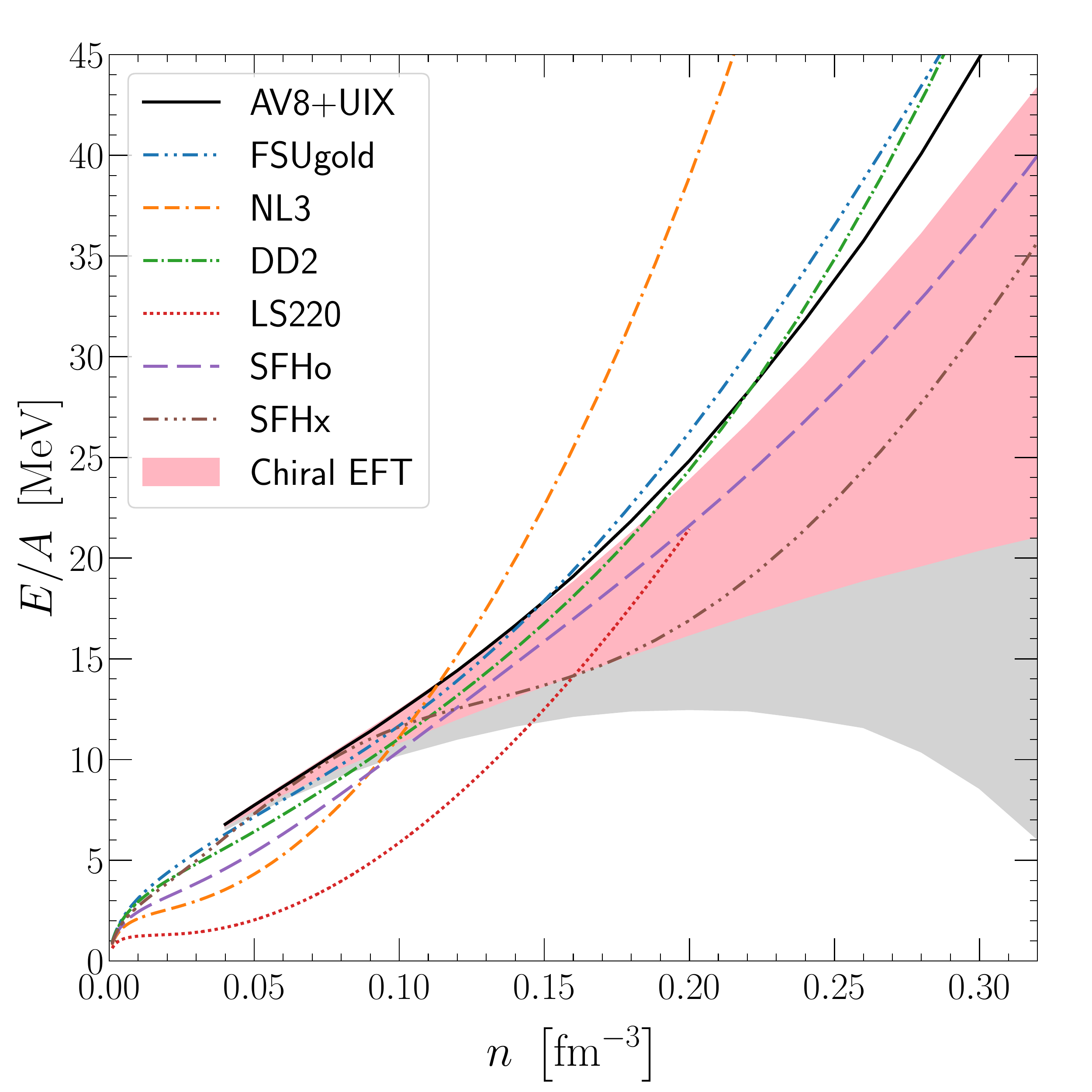}
\caption{The EOS of pure neutron matter calculated using the AFDMC method
with local chiral interactions with a cutoff of $R_0=1.0$ fm (red and gray bands),
and using the AV8'+UIX Hamiltonian (black solid line).
We also compare
the AFDMC results with several EOS that are typically used in astrophysical
simulations. See the text for more details.
\label{fig:PNMEOS}}
\end{center}
\end{figure}

In addition, these interactions give
a reasonable description of PNM~\cite{Lynn:2016,Tews:2018}. We show the PNM EOS calculated using
AFDMC with local chiral interactions and the AV8'+UIX interaction in
Fig.~\ref{fig:PNMEOS}.
The red and gray bands represent AFDMC results using three different
parametrizations of the three-body force constrained in light nuclei,
although they give different results in pure neutron matter due to regulator artifacts; see Ref.~\cite{Lynn:2016} for more details.
In particular, one of them gives negative pressure
in pure neutron matter between $1-2 \rho_0$, and is represented by the gray band in Fig.~\ref{fig:PNMEOS}.
The other two parametrization give instead the red band in the figure, and
can be employed to describe neutron stars~\cite{Tews:2018}.

The AFDMC results are compared to other models that are
commonly used in astrophysical simulations: the Lattimer-Swesty EOS with incompressibility 220~\cite{Lattimer:1991nc}, the TM1, SFHo, and SFHx EOSs~\cite{Hempel}, the FSU and NL3 EOSs~\cite{Shen}, and the DD2 EOS~\cite{Typel}.
We find that AFDMC calculations put strong constraints on the EOS of pure neutron matter.

The AFDMC results for the EOS of PNM can be conveniently parametrized using the functional
\begin{equation}
E_{PNM}(\rho)=a\left(\frac{\rho}{\rho_0}\right)^\alpha+
b\left(\frac{\rho}{\rho_0}\right)^\beta \,,
\label{eq:magic}
\end{equation}
that represents the energy per particle (without the rest
  mass energy) as a function of neutron density. We can define the symmetry energy as
\begin{equation}
E_{\rm sym}(\rho)=E_{\rm PNM}(\rho)-E_{\rm SNM}(\rho) \,,
\end{equation}
where $E_{\rm SNM}(\rho)$ is the EOS of symmetric nuclear matter,
which is not known. However, by requiring that symmetric nuclear matter saturates at an  energy $E_{\rm SNM}(\rho_0)=-16$~MeV, and using
the expression for $E_{\rm PNM}(\rho)$ above, the symmetry energy and slope can be obtained by
\begin{eqnarray}
E_{sym}&=a+b+16 \,,
\nonumber \\
L&=3(a\alpha+b\beta) \,.
\label{eq:Esympar}
\end{eqnarray}
Thus, for $E_{\rm SNM}(\rho)$ we find
\begin{equation}
E_{\rm SNM}(\rho)=E_{\rm PNM}(\rho)-E_{\rm sym}(\rho) \,.
\end{equation}
Assuming a quadratic expansion in the proton fraction $x=\rho_p/\rho$,
the EOS at a finite proton fraction is given by
\begin{eqnarray}
E(\rho,x)&=E_{\rm SNM}(\rho)+E_{sym}(\rho)(1-2x)^2=
\nonumber \\
&=E_{\rm PNM}(\rho)+E_{sym}(\rho)\left[(1-2x)^2-1\right] \,.
\end{eqnarray}
The latter equation gives the expected results for pure neutron matter ($x=0$) and symmetric nuclear matter ($x=1/2$).

A very common parametrization for the symmetry energy is given by
\begin{equation}
E_{sym}(\rho)=C\left(\frac{\rho}{\rho_0}\right)^\gamma \,,
\end{equation}
and using this definition, the slope parameter $L$ is linear with
$C=E_{sym}(\rho_0)$:
\begin{equation}
L=3\rho_0\frac{\partial E_{\rm sym}(\rho)}{\partial\rho}=
3\rho_0\frac{C\gamma}{\rho}\left(\frac{\rho}{\rho_0}\right)^\gamma
\rightarrow L\sim C \,.
\end{equation}
From Eq.~(\ref{eq:Esympar}), we can obtain $C$:
\begin{equation}
C=a+b+16 \,.
\end{equation}
We can put an additional constraint on our simple ansatz by requiring that the pressure of SNM is zero at saturation, i.e.,
\begin{equation}
P=\rho^2\frac{\partial E_{\rm SNM}}{\partial\rho}
|_{\rho=\rho_0}=0 \,.
\end{equation}
This leads to the condition
\begin{equation}
\gamma=\frac{a\alpha+b\beta}{C}=\frac{a\alpha+b\beta}{16+a+b} \,.
\end{equation}
Note that by combining equations, we also find
\begin{equation}
L=3(a\alpha+b\beta) \,.
\end{equation}
With these simple assumptions, the general form of the EOS as a function
of density and proton fraction becomes:
\begin{equation}
E(\rho,x)=E_{\rm PNM}(\rho)+(16+a+b)
\left(\frac{\rho}{\rho_0}\right)^{\frac{a\alpha+b\beta}{16+a+b}}
\left[(1-2x)^2-1\right] \,.
\label{eq:eostotal}
\end{equation}

The parametrizations for selected EOSs are reported in
Table~\ref{tab:eos}, together with the corresponding symmetry energy
and its slope.

\begin{table*}[htbp]
\centering
\begin{tabular}{@{} lcccccc @{}}
\hline
Hamiltonian      & $E_{\rm sym}$ &$L$ & $a$    &  $\alpha$ & $b$ & $\beta$ \\
               & (MeV)  & (MeV) & (MeV)    &        & (MeV)  &   \\
\hline
AV8'                        & 30.5 & 31.3 & 12.7 & 0.49  & 1.78 & 2.26 \\
AV8'+UIX                    & 35.1 & 63.6 & 13.4 & 0.514 & 5.62 & 2.436 \\
N$^2$LO$^{\rm{up}}$ (1.0 fm)   & 34.8 & 56.2 & 13.19 & 0.51  & 5.66 & 2.12  \\
N$^2$LO$^{\rm{low}}$ (1.0 fm)   & 30.2 & 24.4 & 13.87 & 0.59 & 0.36 & -0.05  \\
N$^2$LO$^{\rm{mid}}$ (1.0 fm)  & 32.5 & 40.5 & 13.13 & 0.51 & 3.41 & 1.99\\
\hline
\end{tabular}
\caption{Fitting parameters for the neutron matter EOS defined above. The values
of $E_{sym}$ and $L$ are obtained by fitting the AFDMC results.
The N$^2$LO parameter represent the higher, middle and lower part of the
read band of Fig.~\ref{fig:PNMEOS}.}
\label{tab:eos}
\end{table*}

While $L$ is now strongly constrained by chiral EFT and QMC,
these constraints complement those from X-ray observations of
neutron stars. Ref.~\cite{Steiner12cn} found $43<L<52$~MeV to within
68\% confidence based on an analysis of neutron-star observations
using Eq.~\ref{eq:magic}.

\section{Neutron star properties}

As described above, once the EOS is specified it is easy to calculate
the mass-radius relation of a neutron star. However, since the neutron
star does not consist of pure neutron matter, one has to find ways of
extending the microscopic PNM calculations to neutron-star conditions. In
the following, we will discuss a few possibilities.

\subsection{Effect of leptons}

Typical neutron-star properties can be calculated directly from the
PNM EOS, but such an approach misses the effects of the neutron-star
crust and the remaining protons. Therefore, a more realistic EOS should
contain these effects which can be estimated from the PNM EOS.

\begin{figure}
\begin{center}
\includegraphics[width=0.45\textwidth]{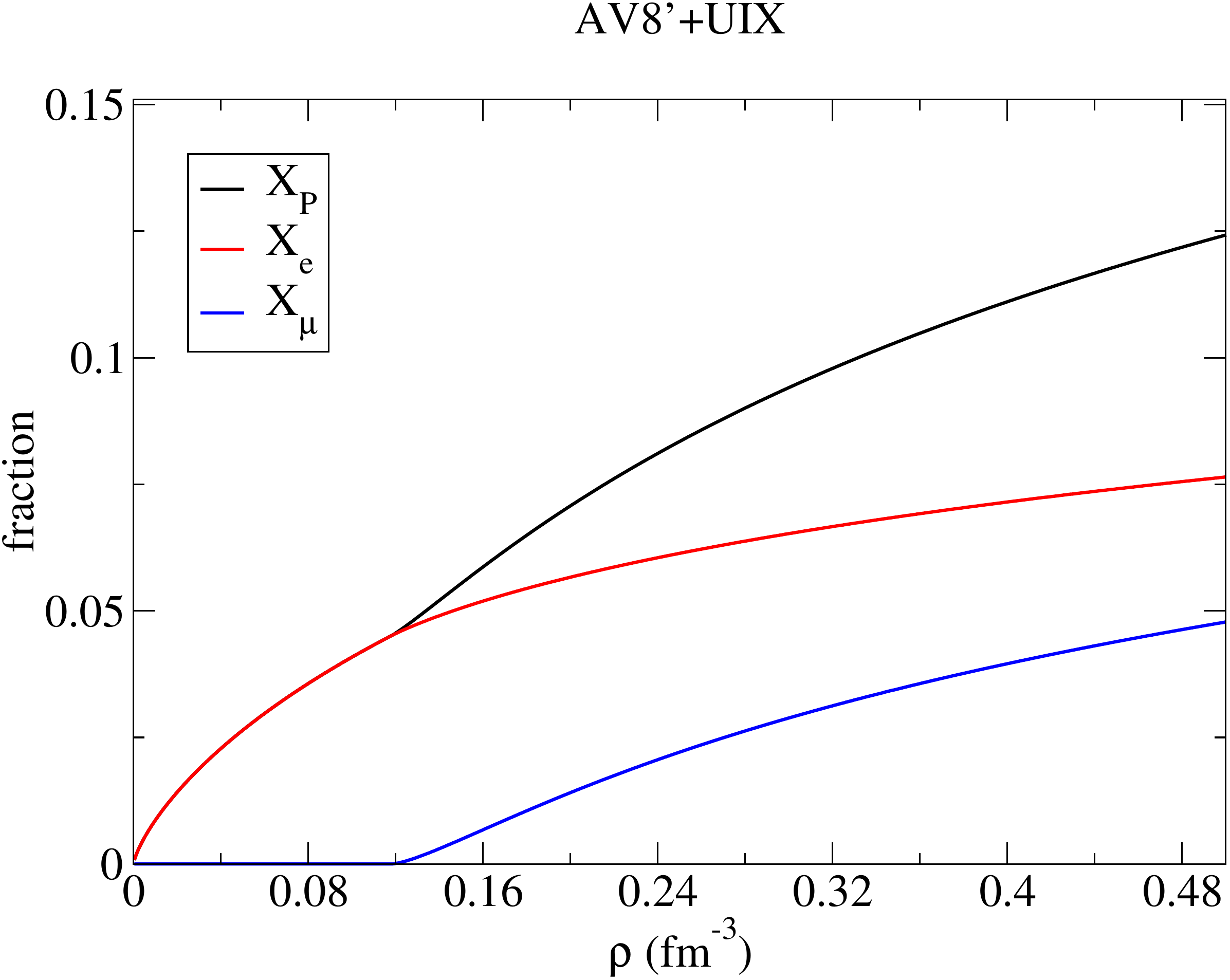}
\includegraphics[width=0.45\textwidth]{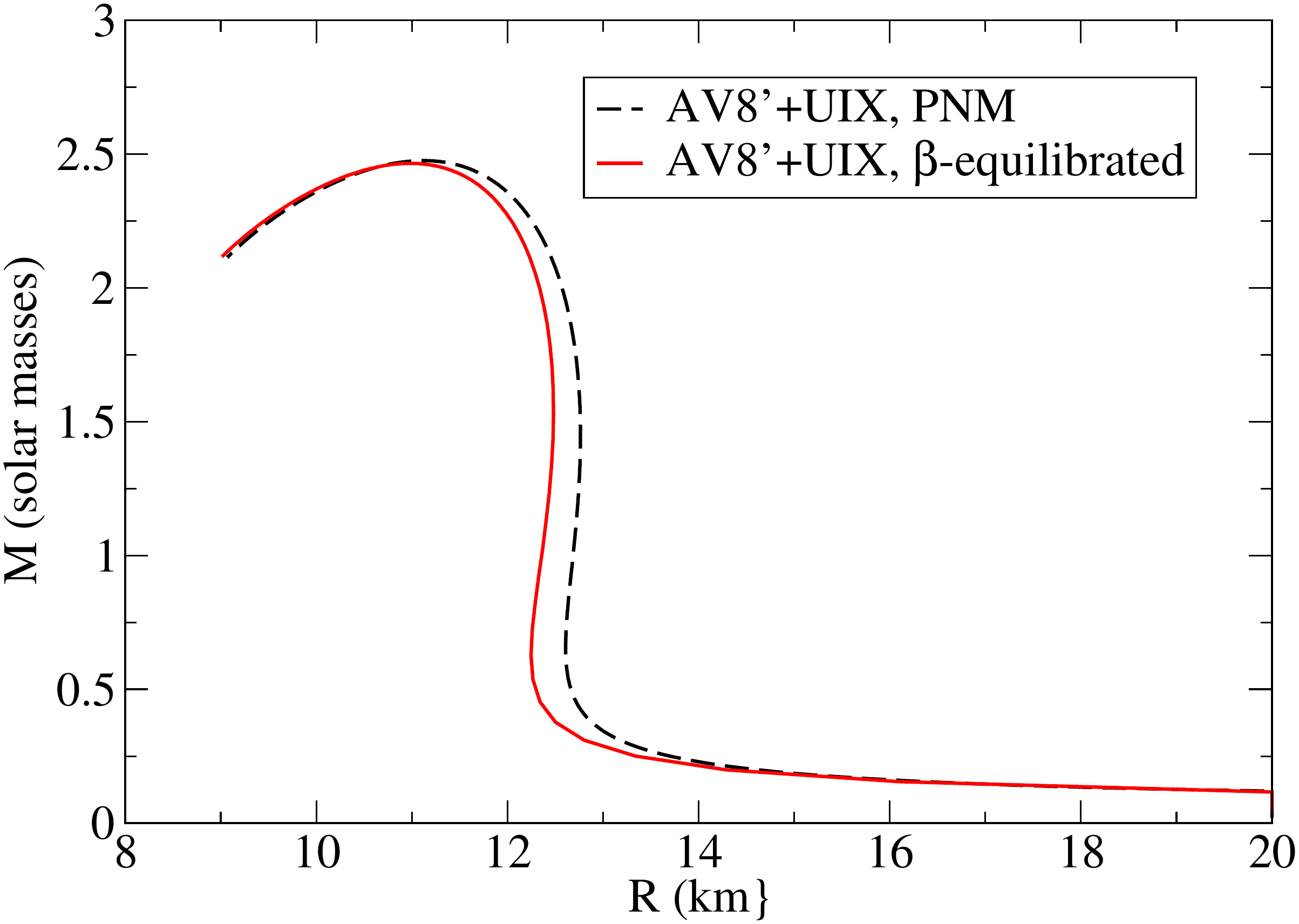}
\caption{Left panel: proton and lepton fractions obtained from
the AFDMC EOS. Right panel: neutron star structure for pure neutron
and $\beta$-equilibrated matter.\label{fig:AV8MR}}
\end{center}
\end{figure}

By starting from Eq.~\ref{eq:eostotal} we can solve for $x(\rho)$ by imposing
$\beta$-equilibrium between neutrons, protons, electrons, and muons.
With
\begin{eqnarray}
\epsilon&=\rho\,[E(\rho,x)+m_n(1-x)+m_p(x)] \,,
\nonumber \\
\mu_y&=\frac{\partial (\rho E)}{\partial\rho_y} \,,
\end{eqnarray}
where $\epsilon$ is the total energy density,
$m_n$ is the neutron mass, $m_p$ is the proton mass,
and $\mu_y$ the chemical potential with $y=n,p$. We can easily
obtain
\begin{equation}
\mu_n-\mu_p=4(1-2x)E_{sym}(\rho) \,.
\end{equation}
Charge neutrality requires that
\begin{equation}
\mu_n-\mu_p=\mu_e=\mu_\mu \\
\end{equation}
We take electrons to be relativistic and degenerate:
\begin{equation}
\mu_e=(m_e^2+\hbar^2 k_F^2)^{1/2}\approx\hbar(3\pi^2\rho x_e)^{1/3} \,,
\end{equation}
and for the muons
\begin{equation}
\mu_\mu=[m_\mu^2+\hbar^2 (3\pi^2\rho x_\mu)^{2/3}]^{1/2} \,.
\end{equation}
We then calculate all the fractions by imposing the charge neutrality,
chemical potential as above, and
\begin{equation}
x=x_e+x_\mu \,.
\end{equation}
Homogeneous matter in $\beta$-equilibrium is a valid model for sufficiently high densities, where nuclei
are not present. Therefore, the AFDMC EOS is used for $\rho\ge\rho_{\rm crust}=0.08$ fm$^{-3}$.
For the low-density EOS, describing the crust of a neutron star, results of earlier works can be used; see e.g. Ref.~\cite{BPS:1971} and~\cite{NegeleVautherin:1973}.
For the AV8'+UIX EOS, we show the MR relations for the PNM EOS and  when we
assume $\beta$ equilibrium in Fig~\ref{fig:AV8MR},
where also the proton and lepton fractions are presented.
We can see that the effect of protons is giving a small correction to the
neutron-star radius given the current uncertainties in
neutron-star observations~\cite{Steiner12cn}.

\subsection{High-density extrapolations}

The approach in the previous section assumes a description in terms of nucleons and leptons to be valid in the whole neutron star. However, while this assumption might be true, the EOS might also explore more extreme density behavior at higher densities, as produced by, e.g.,  strong phase transitions to exotic forms of matter.

\begin{figure}
\begin{center}
\includegraphics[width=0.45\textwidth]{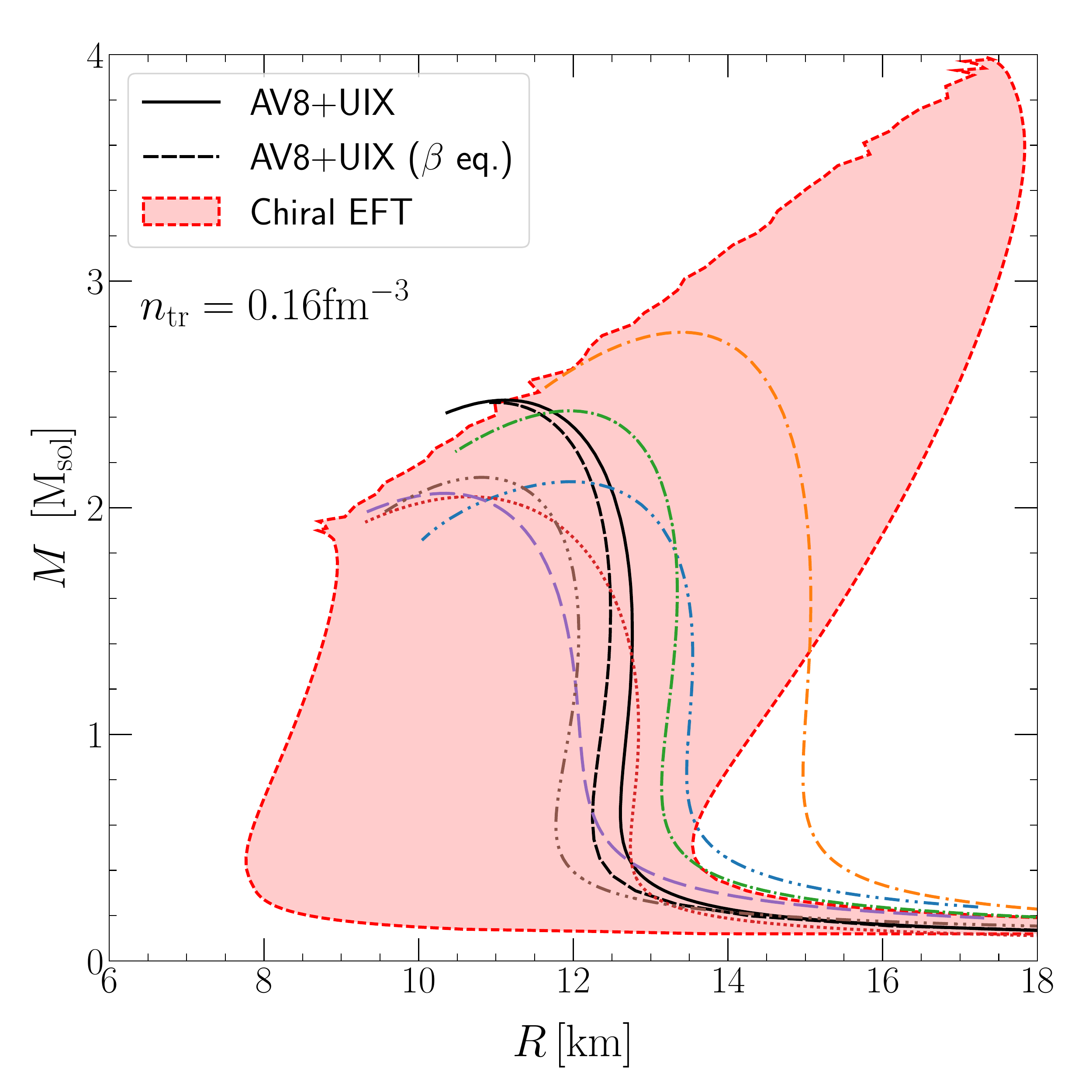}
\includegraphics[width=0.45\textwidth]{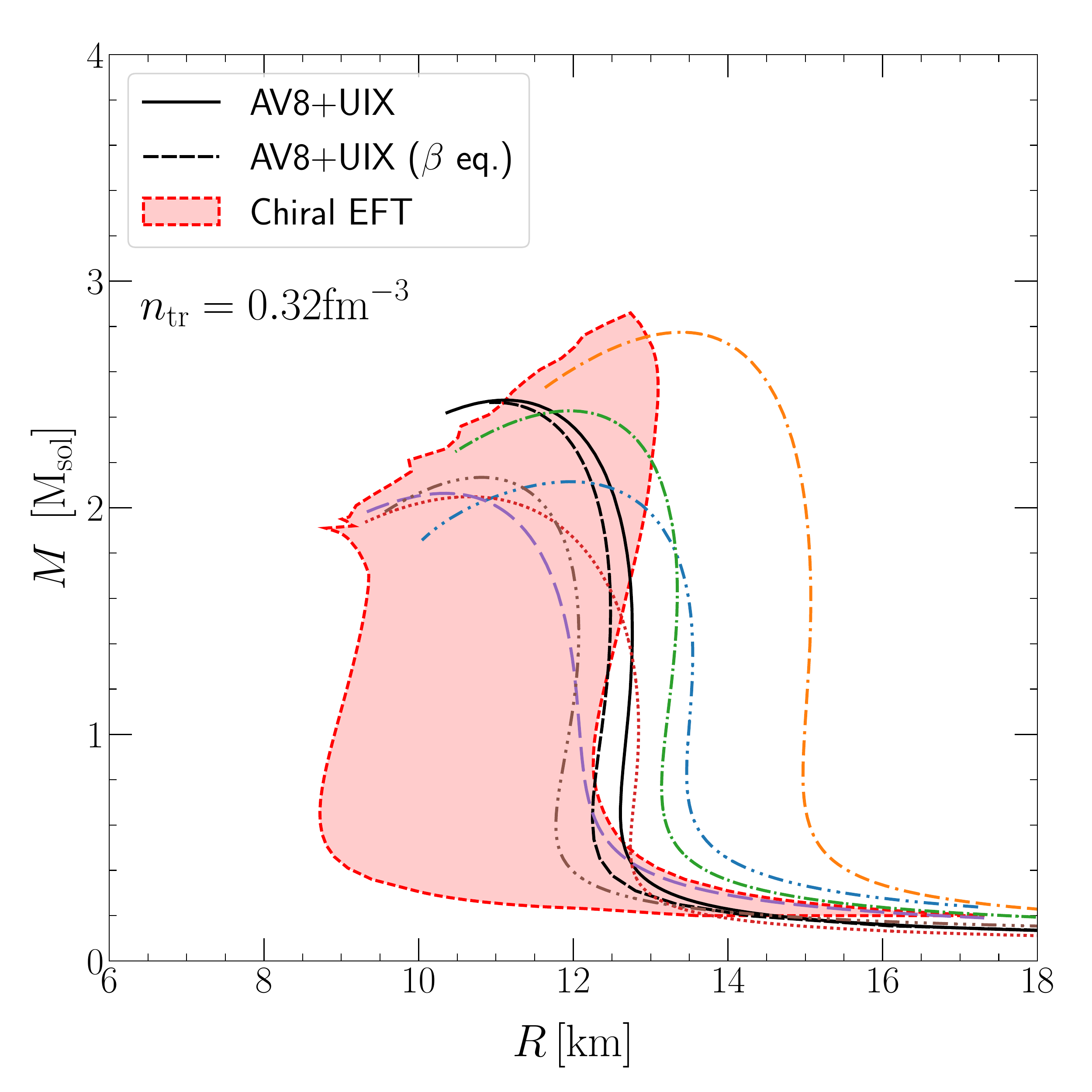}
\caption{General MR extrapolation using the chiral EFT PNM EOS of Fig.~\ref{fig:PNMEOS} up to saturation density (left) or up to two times saturation density (right). We compare with results for the same model EOS as in Fig.~\ref{fig:PNMEOS}.\label{fig:MRchiEFT}}
\end{center}
\end{figure}

To being able to analyze neutron-star properties, general
extrapolations schemes have to be used which are constrained by
nuclear-physics input at low densities as well as observational
constraints. Such general extrapolation schemes can be based on
piecewise polytropes, where polytropes
are line segments in $(\log \epsilon,\log P)$ space; see e.g.,
Refs.~\cite{Read:2008iy,Steiner10te,Hebeler:2013nza}.
However, Ref.~\cite{Steiner13tn} showed that line segments in
$(\epsilon,P)$ space can more easily represent models which have a phase transition
to exotic matter in the neutron-star core.

In contrast, in Refs.~\cite{Tews:2018kmu,Tews:2018chv} another general extrapolation scheme starting from the PNM EOS from chiral EFT was developed, that used the speed of sound, $c_S$ in neutron stars. This scheme was based on the initial work of Refs.~\cite{Alford:2013aca,Bedaque:2014sqa}, but represents an extension of these models by exploring all allowed parameter space  for the speed of sound $c_S$, defined as
\begin{equation}
c_S=\sqrt{\frac{\partial p(\epsilon)}{\partial \epsilon}}\,.
\end{equation}
In particular, models are constrained by the PNM EOS up to a certain density $n_{\rm tr}$ which is varied between 1-2 $\rho_0$.
This PNM EOS is extended to $\beta$-equilibrium and includes a crust as discussed in Ref.~\cite{Tews:2016ofv}. From the resulting neutron-star EOS, the speed of sound is computed up to $n_{\rm tr}$. Beyond this density, many possible paths in the $c_S-n$ plane are explored by randomly sampling several points $c_S^2(n)$ between $n_{\rm tr}$ and 12 $\rho_0$, and connecting them by linear segments. During this procedure, it is enforced that $0\leq c_S \leq c$  and that the resulting EOSs are sufficiently stiff to support a two-solar-mass neutron star~\cite{Demorest:2010,Antoniadis:2013}. For more details on this extrapolation scheme, see Ref.~\cite{Tews:2019cap}.

We show the resulting MR regions in Fig.~\ref{fig:MRchiEFT}, where we compare with the AV8'+UIX result from the previous section as well as the same model EOS of Fig.~\ref{fig:PNMEOS}. We find that chiral EFT input can place strong constraints on the MR relation, ruling out too stiff model EOS, e.g., the NL3 parametrization. In particular, the density range between 1-2 $\rho_0$ is very important to reduce the uncertainty in the MR plane and improved calculations in this density range with smaller uncertainties will be useful to pin down the MR relation of neutron stars.

\section{Connecting the Microphysics with the Multi-messenger Observations}

Multi-messenger astronomy requires a strong foundation of microphysics
in order to fully interpret the observations. The first evidence of
this fact was the observation of both photons and neutrinos from
supernova 1987A. This observation confirmed the basic picture of
stellar evolution, which required significant nuclear and neutrino
physics input\cite[e.g.][]{burrows:87b,bruenn:87,
  hirata:87,arnett:89,Langanke03,pagliaroli:09b}. More recently, the
first gravitational-wave observation of a binary neutron-star merger,
GW170817, has further highlighted the tight connection between
microphysics, including the nuclear EOS, and multi-messenger astronomy
observations \cite[e.g.][]{ligo:17nsns, ligo:17nsns_dyn_ej}.

Neutron-star mergers have long been thought to be the progenitors of short
gamma-ray bursts (GRBs) and a significant, if not dominant, site of r-process
nucleosynthesis \cite[e.g.][]{lattimer:76, mochkovitch:93, janka:99,
freiburghaus:99, wlee:07, nakar:07a, gehrels:09, fong:13, rosswog:15b}. This
basic picture was confirmed by GW170817, which was observed in coincidence with
a short GRB (GRB170817A) from the same location in the sky
\cite[e.g.][]{ligo:17nsnsmm, goldstein:17, savchenko:17}. Subsequent followup
observations with instruments and telescopes spanning the entire
electro-magnetic spectrum revealed a kilonova \cite[e.g.][]{ligo:17nsnsmm,
cowperthwaite:17, tanvir:17, tanaka:17, coulter:17, evans:17}. Kilonovae are
rapidly fading (timescale of a few weeks) optical and infrared transients that
are powered by the radioactive decay of the newly synthesized r-process elements
\cite[e.g.][]{li:98, lippuner:15, fernandez:16, metzger:17a, kasen:17}. They are
the smoking gun signatures of r-process nucleosynthesis events.

\begin{figure}[t]
\begin{center}
\includegraphics[width=0.8\textwidth]{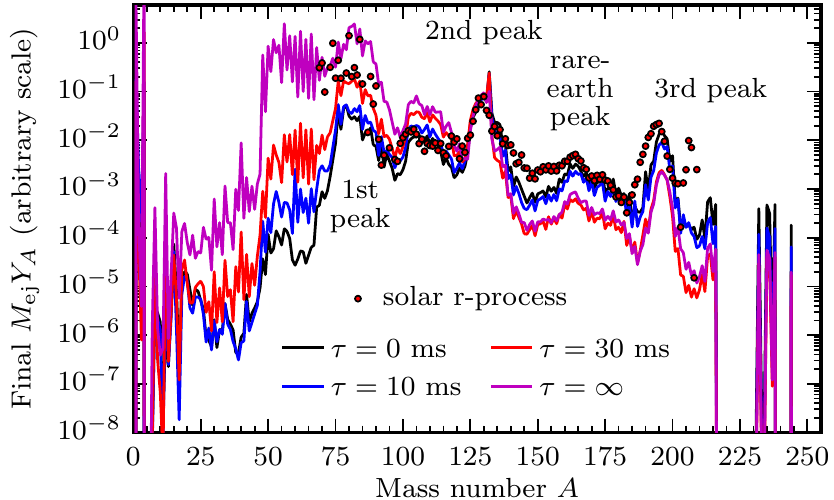}
\caption{Final abundances times the total ejecta mass $M_\mathrm{ej}$ of
accretion disk outflows around a hypermassive neutron star (HMNS) for different
lifetimes $\tau$ of the HMNS. If the HMNS collapses to a black hole within
10~ms, the resulting nucleosynthesis pattern matches the solar pattern quite
well. But if the lifetime is $\tau \gtrsim 30$~ms (including if the HMNS never
collapses), the rare-earth and 3rd r-process peaks are suppressed with respect
to the solar pattern. The dots show the solar r-process pattern from
\cite{arnould:07}. This is an adapted version of Fig.~4 in \cite{lippuner:17a},
see that reference for details.
\label{fig:hmns-lifetime}}
\end{center}
\end{figure}

Even though GW170817 confirmed some aspects of the basic picture of neutron-star
mergers and the role they play in r-process nucleosynthesis, many details remain
unresolved. Over the coming years, LIGO/VIRGO will observe more neutron-star
mergers and we should be able to find the associated kilonovae for at least some
of them \cite[e.g.][]{ligo:18_prospects, scolnic:18, metzger:17b, rosswog:17,
doctor:17}. A detailed understanding of the microphysics will be crucial in
accurately modelling the rich multi-messenger observational data from these
events to fully understand them and extract system parameters that are not
directly observable. Specifically, microphysics and the nuclear EOS directly
impact the amount and possibly morphology of the mass ejecta and accretion disk
\cite[e.g.][]{duez:10a, hotokezaka:11, deaton:13, bauswein:14, palenzuela:15,
kawaguchi:15, radice:16b, foucart:17a, ligo:17nsns_dyn_ej, radice:18,
kyutoku:18} that forms around the compact central remnant. They also determine
how long a hypermassive neutron star (HMNS) could live, if one forms, and this lifetime
affects the amount of ejecta blown away from the disk and its neutron-richness,
which in turn determines what elements the r-process can synthesize
\cite[e.g.][]{martin:15, lippuner:17a}. An example of the effect of the lifetime
of the HMNS is shown in Fig.~\ref{fig:hmns-lifetime}.
Neutrino interactions in general are
very important to determine the composition of the ejecta and subsequent
nucleosynthesis \citep[e.g.][]{wanajo:14, metzger:14, sekiguchi:15, foucart:15,
goriely:15, foucart:16a, roberts:17b, siegel:17}. All of these properties
affected by microphysics directly influence the kilonova lightcurve,
nucleosynthetic yields, and possibly the gravitational wave signal. Therefore,
we need to get the microphysics right in order to draw meaningful conclusions
from neutron-star merger and kilonova observations about their intrinsic
properties and how they enrich the galaxy with heavy elements
\cite[e.g.][]{cote:18}.

\subsection{Constraining the Microphysics of Merger Simulations
  with GW170817}

Neutron-star merger simulations require a three-dimensional EOS table:
a description of several thermodynamic quantities as a function of the
baryon density, $\rho$, electron fraction, $Y_e$, and the temperature,
$T$. Until recently, only about a dozen of these EOS tables was
available. These EOS tables explore only a tiny fraction of the large
space of EOSs which appear to be compatible with our current knowledge
of the nucleon-nucleon interaction. For a review, see Ref.~\cite{Oertel17eo}.

This state of affairs changed when Ref.~\cite{Schneider17} released
an open-source code for EOS tables built upon the Skyrme interaction.
This code allows one to fully explore a large space of EOSs. In
particular, a merger simulation may systematically probe the
sensitivity of the observables generated by the merger simulation to
the parameters in the Skyrme interaction. This code, however,
is limited by the applicability of the Skyrme interaction.

In Ref.~\cite{Du18hd} we presented a new class of phenomenological EOS
for homogeneous nucleonic matter. The EOS is constructed to
simultaneously match (i) second order virial expansion coefficients
from nucleon-nucleon scattering phase shifts at high temperature and
low density, (ii) experimental results of nuclear mass and radii,
(iii) QMC calculations for neutron matter at saturation density, (iv)
astrophysical observational measurements on neutrons star radii, (v)
theoretical calculations with chiral perturbation field theory at
finite temperature near the saturation density. It allows for
computing the variation in the thermodynamic quantities based on the
uncertainties of nuclear interaction.

Our free energy per particle  can be written as
\begin{equation}
  F_{\mathrm{np}}(\rho,x,T) =F_{\mathrm{virial}}(\rho,x,T) g  + F_{\mathrm{deg}}(\rho,x,T) (1-g)\,,
\end{equation}
where $x$ is the number of protons per baryon (assumed here to be
equal to $Y_e$), $f_{\rm virial}$ is the virial free energy
contribution~\cite{Horowitz06} and $f_{\rm deg}$ is the free energy
for degenerate matter. The function is defined by
\begin{equation}
g=1/(1+3 z_n^2 + 3 z_p^2)\,,
\end{equation}
where $z_i\equiv \exp(\mu_i/T)$ are the fugacities. Since the virial
expansion is valid when $z_n^2, z_p^2 \ll 1$, this functional form
gives $g \approx 1$, thus the dominant contribution is from $F_{\rm
  virial}$. Otherwise, if $z_n$ or $z_p$ are large, $F_{\rm np}
\approx F_{\rm deg}$. The free energy per particle of degenerate
matter is further defined assuming quadratic expansion
\begin{eqnarray}
F_{\mathrm{deg}}(\rho,x,T) &=& F_{\mathrm{Skyrme}}(\rho,x=1/2,T=0)
+ \delta^2 {E}_{\mathrm{sym}}(\rho) \\ \nonumber
&& + F_{\rm hot} (\rho,x,T) - F_{\rm hot} (\rho,x=0,T)
\end{eqnarray}
with $\delta = 1- 2 x $.
The 1000 parameter set of Skyrme model was chosen from UNEDF collaboration fitted to several nuclear mass, charge radii and pairing energies using Bayesian inference \cite{Kortelainen14}.
The $F_{\rm hot}$ is finite temperature results based on Kohn-Luttinger-Ward pertubation series fitted to Skyrme functional form \cite{Wellenhofer+14,Wellenhofer15to}.
 The symmetry energy is defined by
\begin{eqnarray}
E_{\rm sym} = h(\rho) E_{\rm PNM} +[1-h(\rho)] E_{\rm NS} (\rho) - F_{\rm Skyrme}
(\rho,x=1/2,T=0)\,,
\end{eqnarray}
where we interpolate between Eq.~(\ref{eq:magic}) near saturation
density $\rho_0$ and a polynomial fit to neutron star observational
data \cite{Steiner10te} above $2 \rho_0$ using a function $h$, defined as
\begin{equation}
h= \frac{1}{1+\exp[\gamma (\rho - 3/2 \rho_0)]}\,,
\end{equation}
where $\gamma=20.0 \, {\rm fm}^3$. At zero temperature, the EOS can be
compared to Eq.~(\ref{eq:eostotal}).

\begin{figure}
\begin{center}
\includegraphics[width=0.49\textwidth]{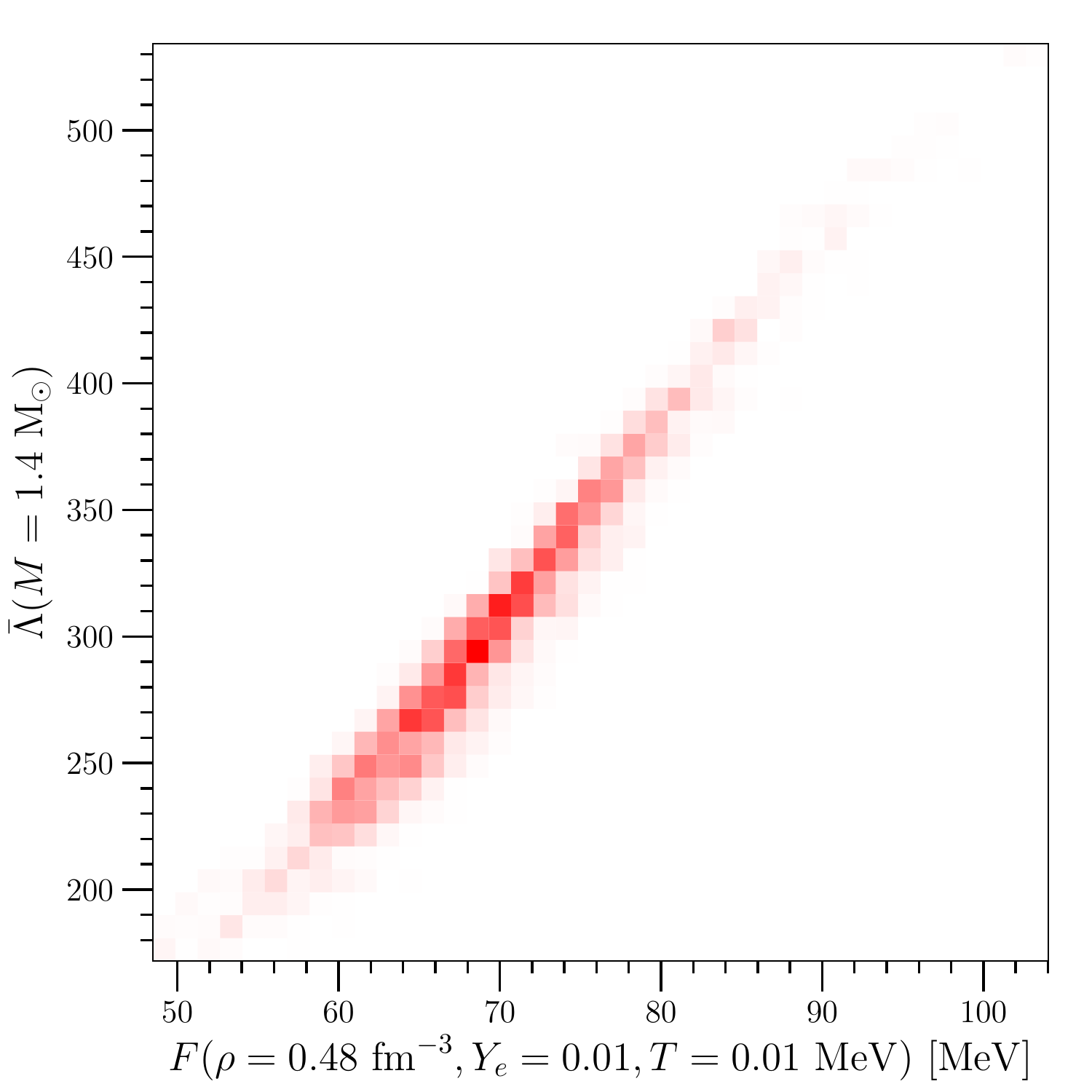}
\includegraphics[width=0.49\textwidth]{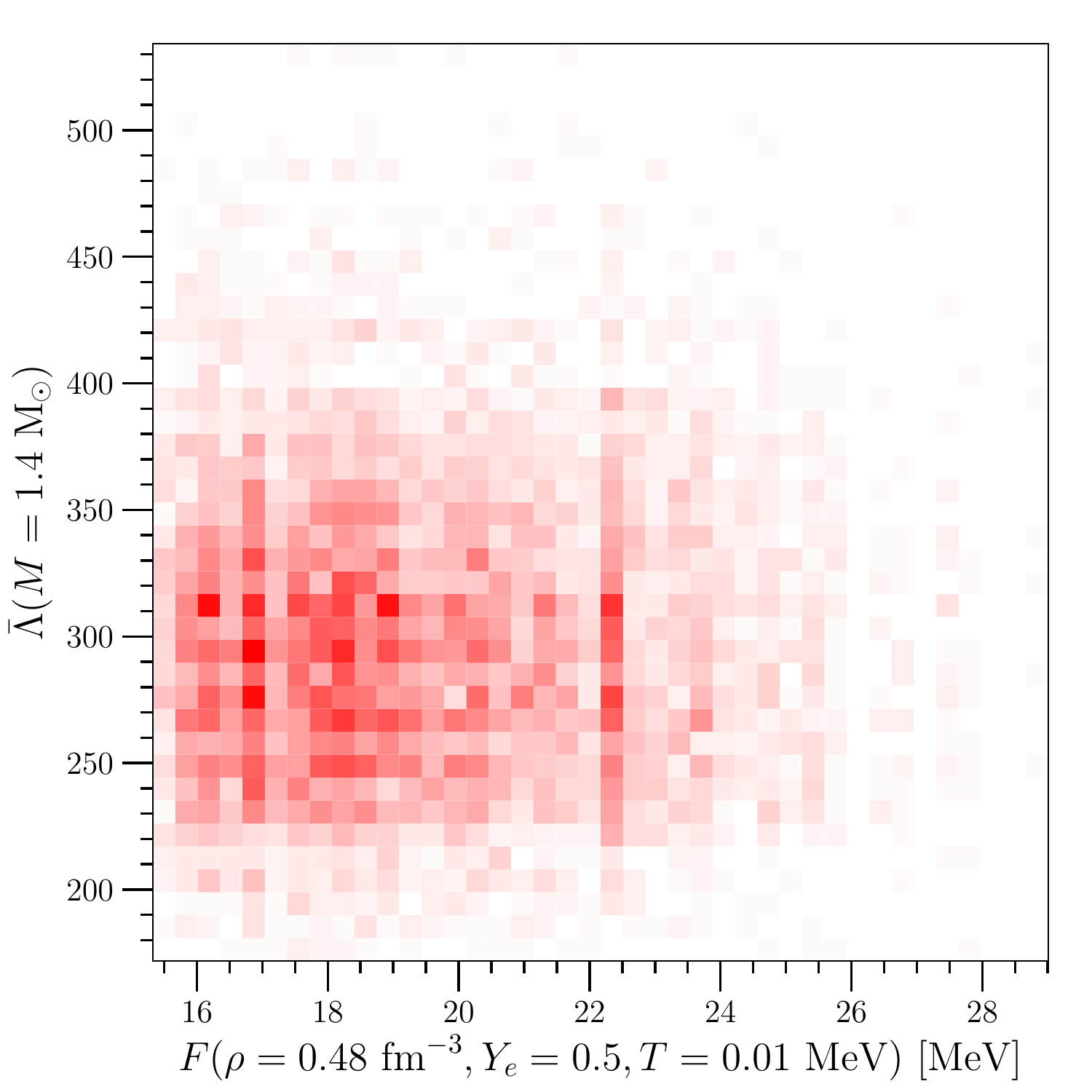}
\caption{Two-dimensional histograms of the tidal deformability versus
  the free energy per particle from the DSH formalism~\cite{Du18hd} at
  $\rho=0.48~\mathrm{fm}^{-3}$ in either neutron-rich matter (left
  panel) or isospin-symmetric matter (right panel). The distribution
  of tidal deformabilities already accounts for the constraints due to
  neutron-star mass and radius constraints from X-ray
  observations~\cite{Steiner15un}. The left panel shows that a
  constraint on the tidal deformability has a strong impact on
  constraining the EOS of neutron-rich matter. However, the right
  panel shows that GW observations provide almost no constraint on the
  EOS of symmetric nuclear matter. This demonstrates the role of the nuclear
  symmetry at high densities: while neutron-star observations
  constrain neutron-rich matter, they do not constrain the symmetry
  energy at high densities unless one assumes that the EOS of nuclear
  matter is otherwise fixed.}
\label{fig:dsh}
\end{center}
\end{figure}

The advantage of the formalism above is that it allows us not
  only to compute the EOS of homogeneous nucleonic matter over the
  full range of densities, electron fractions, and temperatures, but
  it allows us to describe the probability distribution of the EOS.
  Formulating the EOS in this way allows us to easily determine the
  impact that observations might have on the EOS. This is demonstrated
  in Fig.~\ref{fig:dsh} where the correlation between the free energy
  per baryon at $\rho = 3 \rho_0$ and the tidal deformability of a 1.4
  $\mathrm{M}_{\odot}$ neutron star is shown.  The tidal polarizability describes how a neutron star deforms under an external gravitational field, as produced by a companion star. It is given by
\begin{equation}
\Lambda=\frac23 k_2 \left(\frac{c^2}{G} \frac{R}{M}\right)^5\,,
\end{equation}
with the tidal Love number $k_2$ which has to be solved together with the TOV equations; see, e.g., Ref.~\cite{Damour:2009vw}.
  As is expected, the free
  energy of neutron-rich matter is strongly corrlated with the tidal
  deformability, but the uncertainty in the free energy of
  isospin-symmetric matter is not impacted by constraints on the tidal
  deformability.

\subsection{EOS constraints from tidal polarizabilities}

\begin{figure}[t]
\begin{center}
\includegraphics[width=1.0\textwidth]{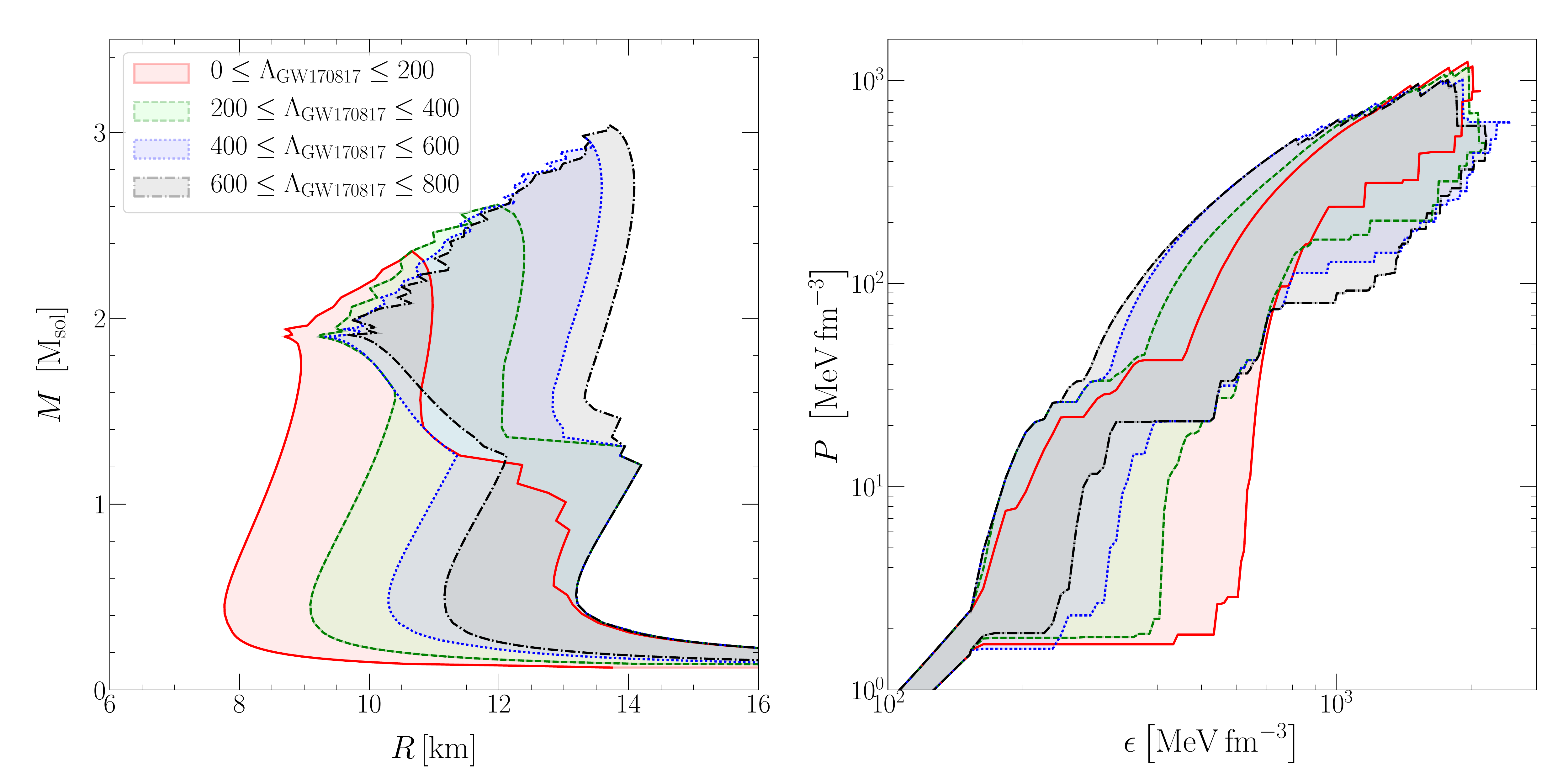}
\caption{The MR relation and EOS using chiral EFT input up to $n_{\rm tr}=\rho_0$, if the tidal polarizability of GW170817 could be constrained to lie in the respective ranges.
\label{fig:MRLambda}}
\end{center}
\end{figure}

The analysis of the gravitational-wave signal from the inspiral phase of two merging neutron stars allows to place constraints on properties of the binary system, e.g., the chirp mass or mass ratio, as well as on properties of individual neutron stars. For a neutron-star binary, the binary tidal polarizability can be defined as
\begin{equation}
\tilde{\Lambda}=\frac{16}{13}\left(\frac{(m_1+12 m_2)m_1^4\Lambda_1}{m_{\rm tot}^5}+\frac{(m_2+12 m_1)m_2^4\Lambda_3}{m_{\rm tot}^5}   \right)\,.
\end{equation}
This quantity, $\tilde{\Lambda}$, can be constrained from the GW signal of a NS merger. For the first NS merger observed, GW170817, this tidal polarizability originally was constrained to be $\tilde{\Lambda}\leq 800$ at a 90\% confidence level~\cite{AbbottPRL:2017} but later modified after several reanalyses~\cite{De:2018uhw, Abbott:2018wiz} to values around $70\leq \tilde{\Lambda}\leq 700$.

Qualitatively, neutron stars with larger radii can exhibit stronger deformations than small NS and, therefore, also have larger tidal polarizabilities. Hence, the tidal polarizability is closely related to the structure of a NS and can help to place constraints to the EOS. Using the chiral EFT results discussed in Sec.~\ref{sec:EOS} up to saturation density (left panel of Fig.~\ref{fig:MRchiEFT}),
in Fig.~\ref{fig:MRLambda} we show the resulting MR and EOS envelopes if $\tilde{\Lambda}$ would be constrained to the 4 specified ranges, stretching from $0\leq\tilde{\Lambda}\leq 800$. While the observation GW170817 basically allows for a total radius range of $8.4-13.6$ km for a typical $1.4 M_{\odot}$ neutron star, we see that more precise observations in future might help to drastically constrain the radii of neutron stars and, therefore, the EOS. For example, constraining the tidal polarizability to lie in an interval with a width $\Delta \tilde{\Lambda}\approx 200$, reduces the radius uncertainty to $\approx 2$ km for a typical neutron star.

\subsection{Bayesian Inference: Using Neutron Star Observations to
Determine the EOS}

A method to directly determine the EOS from neutron-star mass and
radius observations was first described in Ref.~\cite{Ozel09rt}.
Shortly thereafter, Ref.~\cite{Steiner10te} presented an alternate
method using Bayesian inference. Ref.~\cite{Steiner10te} also obtained
the first quantitative results, obtaining uncertainties in the EOS
around 30\% just above the nuclear saturation density. Bayesian
inference is still the tool of choice for extracting the EOS, although
many works employ simple Monte Carlo methods which are identical to
Bayesian results with trivial priors.

One critical question (which may be soon answered) is whether or not
the EOS contains a strong phase transition: a region in the EOS where
the derivative of the pressure with respect to the energy density
(identical to the speed of sound) is nearly zero. Because the pressure
must be continuous in the star, a strong phase transition creates a
thin region in the neutron star where the energy density increases
strongly with decreasing radius. Strong phase transitions of this type
can result in mass-radius curves which have multiple branches, as
described in Ref.~\cite{Alford13gc} and recently explored in
Ref.~\cite{Han18td}.

Below, we show that multiple branches in the mass-radius curve cannot
be easily handled in the Bayesian methods which have been previously
used in the literature and we show how this difficulty can be fixed.
The traditional method for extracting the mass-radius curve from $N$
neutron star mass and radius determinations (each expressed in the
form of a two-dimensional probability distribution ${\cal D}(R,M)$) is
to use a likelihood function of the form
\begin{equation}
  {\cal L}(\{p_j\}) = \int \prod_{i=1}^{N}dM_i {\cal
    D}_i[R(M_i,\{p_j\}),M_i] \, ,
  \label{eq:like}
\end{equation}
where $\{p_j\}$ are the parameters of the EOS and the TOV equations
are solved to obtain the function $R(M_i,\{p_j\})$. (This likelihood
is analogous to Eq. (15) of Ref.~\cite{Ozel09rt} and Eqs. (29) and (31) of
Ref.~\cite{Steiner10te}, but the formalism in these two works is a bit
different, see Ref.~\cite{Riley18op} for an interesting discussion
comparing the two works.) The critical feature of these likelihoods is
that they are both expressed as integrals over the gravitational mass.
It is thus clear that these likelihood functions are identically zero
for horizontal mass-radius curves.

In writing Eq.~(\ref{eq:like}) we specifically made a choice to
parametrize the M-R curve with the gravitational mass rather than
the radius. Alternatively, we could have chosen the radius
instead
\begin{equation}
  {\cal L}(\{p_j\}) = \int \prod_{i=1}^{N}dR_i {\cal
    D}_i[R_i,M(R_i,\{p_j\})] \, ,
  \label{eq:rlike}
\end{equation}
where the TOV equations now provide the function $M(R_i,\{p_j\})$.
This likelihood is not equivalent to Eq.~(\ref{eq:like}); it results in
different posterior distributions and, given an EOS parametrization,
will be maximized at a different location in parameter space. To
see this, note that Eq.~(\ref{eq:like}) can be rewritten as
\begin{equation}
  {\cal L}(\{p_j\}) = \int \prod_{i=1}^{N}dR_i {\cal
    D}_i[R_i,M(R_i,\{p_j\})]
  \sqrt{1+\left(\frac{dM_i}{dR_i}\right)_{\{p_j\}}}
  \label{eq:mliket}
\end{equation}
which is not equal Eq.~(\ref{eq:rlike}).

Choosing either Eq.~(\ref{eq:like}) or Eq.~(\ref{eq:rlike}) is
further problematic
because M-R curves like those from EOSs with strong phase transitions
can result in two (or even three) configurations with the same
gravitational mass yet different radii. The only way to use the
likelihood in Eq.~(\ref{eq:like}) is to break up the
integral into different parts for each branch of the M-R curve
\begin{eqnarray}
  {\cal L}(\{p_j\}) &=&
  \int_{M_0}^{M_1} \prod_{i=1}^{N}dM_i {\cal D}_i[R(M_i,\{p_j\}),M_i]
  \nonumber \\
  && + \int_{M_2}^{M_3} \prod_{i=1}^{N}dM_i {\cal D}_i[R(M_i,\{p_j\}),M_i]
  \label{eq:break}
\end{eqnarray}
where $M_0<M_1$ and $M_2<M_3$ but $M_2<M_1$. This form of the
likelihood makes the numerical evaluation of the likelihood more
awkward. One possible resolution is to use the central neutron
star pressure, $P$ instead of the gravitational mass $M$ or radius
$R$, i.e.
\begin{equation}
  {\cal L}(\{p_j\}) = \int \prod_{i=1}^{N}dP_i
  {\cal D}_i[R(P_i,\{p_j\}),M(P_i,\{p_j\})] \, .
  \label{eq:plike}
\end{equation}
This integral is simpler because the functions $R(P_i,\{p_j\})$ and
$M(P_i,\{p_j\})$ are never multiply-valued and the pressure is
continuous within the star. This form has the additional advantage
that neither ``horizontal'' nor ``vertical'' mass-radius curves are
arbitrarily given zero likelihood, though this choice is still part of
the prior distribution. Note that choosing the central energy density,
instead of the central pressure, as suggested by Eq. (14) in
Ref.~\cite{Ozel16td} also leads to multiple integrals as in
Eq.~(\ref{eq:break}) when strong phase transitions create regions
where $dP/d\varepsilon=0$.

Ref.~\cite{Steiner18ta} recently clarified why the ambiguity between
Eqs.~(\ref{eq:like}) and (\ref{eq:rlike}) arises. It originates in the fact
that we are attempting to match a one-dimensional model curve (the
size of the EOS parameter space is not important for this discussion)
to a two-dimensional data set ${\cal D}(R,M)$. In the language of
differential geometry, we are embedding the curve into the data space
and the ambiguity we found above is related to the choice of how we
perform this embedding. The general form requires a metric ($g$) to
specify how distances are defined
\begin{eqnarray}
  {\cal L}(\{p_j\}) &=& \int \prod_{i=1}^{N}d\lambda_i
  {\cal D}_i[R(\lambda_i,\{p_j\}),M_i(\lambda_i,\{p_j\})]
  \nonumber \\ && \times \left[g_{jk}
  \left(\frac{dX_j}{d\lambda_i}\right)_{\{p_j\}}
    \left(\frac{dX_k}{d\lambda_i}\right)_{\{p_j\}}\right] \, .
  \label{eq:llike}
\end{eqnarray}
where $j$ and $k$ range from 1 to 2 and $X_1 \equiv M$ and $X_2 \equiv
R$ and for simplicity we can assume the metric is independent of $i$.
The ambiguity in choosing the likelihood function is now explicit in
the choice of the metric, $g_{jk}$. In the context of Bayesian
inference, this metric choice is to be regarded as part of the prior
probability. Choosing $\lambda_i=P_i$ and a trivial metric
$g_{jk}=\delta_{jk}$ gives Eq.~(\ref{eq:plike}). It is clear, however,
that other choices of the metric (the elements may depend on
$\lambda$ or $X$) will give different results.

The implication, in the context of future gravitational wave
observations of neutron star mergers, is that unambiguous
determinations of the EOS will have to wait until sufficient data is
obtained as to make the posterior distributions of quantities of
interest (such as EOS parameters) are not strongly dependent on the
choice of prior probability. This prior probability
includes the EOS parametrization
and the prior probabilities for the EOS parameters~\cite{Steiner16ns}
as well as the prior choice of metric in Eq.~(\ref{eq:llike}).

\section{Summary}

In this paper, we have reviewed current calculations of the EOS of neutron-star matter starting from AFDMC calculations of pure neutron matter using interactions from chiral EFT. We have presented the EOS of PNM with theoretical uncertainties and explained, how to extend these PNM calculations to neutron-star conditions. We then used this to explore current theoretical uncertainties for the mass-radius relation of neutron stars.

We have then discussed how this microphysics can be connected to multimessenger observations of neutron-star mergers, using several EOS models or Bayesian inference.

Neutron-star mergers offer an ideal way to constrain the EOS of strongly interacting matter, and thus, nuclear interactions.  To pin down the nuclear interaction, theorists working on a solid theoretical description of microphysics, computational astrophysicists who simulate neutron-star mergers and supernovae, and observers have to work hand-in-hand to reliably extract constraints from future merger observations.

\section*{Acknowledgments}

The work of S.G. and I.T. was supported by
the U.S. DOE under contract DE-AC52-06NA25396 and by the LANL LDRD program. The
work of S.G. was also supported by the NUCLEI SciDAC program and by the DOE
Early Career Research Program.
the work of JL was supported by the Laboratory Directed Research and Development
program of Los Alamos National Laboratory under project number 20190021DR.
The work of XD and AWS was supported by DOE SciDAC grant DE-SC0018232.
The work of MA and AWS was supported by NSF grant PHY 1554876.
This work was supported by the US Department of Energy through the Los Alamos
National Laboratory and used resources provided by the Los Alamos National
Laboratory Institutional Computing Program.
Los Alamos National Laboratory is operated by Triad National Security, LLC, for
the National Nuclear Security Administration of U.S. Department of Energy
(Contract No.\ 89233218CNA000001).
We also used resources provided by NERSC,
which is supported by the US DOE under Contract DE-AC02-05CH11231, and by the J\"ulich Supercomputing Center.

Figs. 2 and 3 are open source and available at \href{https://github.com/awsteiner/nstar-plot}{https://github.com/awsteiner/nstar-plot}.
\\

\bibliographystyle{iopart-num}

\begin{thebibliography}{100}
\expandafter\ifx\csname url\endcsname\relax
  \def\url#1{{\tt #1}}\fi
\expandafter\ifx\csname urlprefix\endcsname\relax\def\urlprefix{URL }\fi
\providecommand{\eprint}[2][]{\url{#2}}

\bibitem{Burbidge57}
{Burbidge} E~M, {Burbidge} G~R, {Fowler} W~A and {Hoyle} F 1957 {\em Rev. Mod.
  Phys.\/} {\bf 29} 547--650
  \urlprefix\url{https://doi.org/10.1103/RevModPhys.29.547}

\bibitem{Cameron59}
{Cameron} A~G 1959 {\em \apj\/} {\bf 130} 884
  \urlprefix\url{https://doi.org/10.1086/146780}

\bibitem{Demorest:2010}
Demorest P~B, Pennucci T, Ransom S~M, Roberts M~S~E and Hessels J~W~T 2010 {\em
  Nature\/} {\bf 467} 1081--1083

\bibitem{Antoniadis:2013}
Antoniadis J, Freire P~C~C, Wex N, Tauris T~M, Lynch R~S, van Kerkwijk M~H,
  Kramer M, Bassa C, Dhillon V~S, Driebe T, Hessels J~W~T, Kaspi V~M,
  Kondratiev V~I, Langer N, Marsh T~R, McLaughlin M~A, Pennucci T~T, Ransom
  S~M, Stairs I~H, van Leeuwen J, Verbiest J~P~W and Whelan D~G 2013 {\em
  Science\/} {\bf 340} 1233232
  \urlprefix\url{http://www.sciencemag.org/content/340/6131/1233232.abstract}

\bibitem{Fonseca16}
{Fonseca} E, {Pennucci} T~T, {Ellis} J~A, {Stairs} I~H, {Nice} D~J, {Ransom}
  S~M, {Demorest} P~B, {Arzoumanian} Z, {Crowter} K, {Dolch} T, {Ferdman} R~D,
  {Gonzalez} M~E, {Jones} G, {Jones} M~L, {Lam} M~T, {Levin} L, {McLaughlin}
  M~A, {Stovall} K, {Swiggum} J~K and {Zhu} W 2016 {\em \apj\/} {\bf 832} 167

\bibitem{Ozel10am}
Ozel F, Baym G and Guver T 2010 {\em Phys. Rev.\/} {\bf D82} 101301
  \urlprefix\url{https://doi.org/10.1103/PhysRevD.82.101301}

\bibitem{Steiner10te}
Steiner A~W, Lattimer J~M and {Brown} E~F 2010 {\em Astrophys. J.\/} {\bf 722}
  33--54 \urlprefix\url{https://doi.org/10.1088/0004-637X/722/1/33}

\bibitem{Ozel16td}
Ozel F, Psaltis D, Guver T, Baym G, Heinke C and Guillot S 2016 {\em Astrophys.
  J.\/} {\bf 820} 28 \urlprefix\url{https://doi.org/10.3847/0004-637X/820/1/28}

\bibitem{Steiner18ct}
Steiner A~W, Heinke C~O, Bogdanov S, Li C, Ho W~C~G, Bahramian A and Han S 2018
  {\em Mon. Not. Roy. Astron. Soc.\/} {\bf 476} 421
  \urlprefix\url{https://doi.org/10.1093/mnras/sty215}

\bibitem{Nattila17ns}
N\"{a}ttil\"{a} J, Miller M~C, Steiner A~W, Kajava J~J~E, Suleimanov V~F and
  Poutanen J 2017 {\em Astron. and Astrophys.\/} {\bf 608} A31
  \urlprefix\url{https://doi.org/10.1051/0004-6361/201731082}

\bibitem{ligo:17nsns}
{Abbott} B~P, {Abbott} R, {Abbott} T~D, {Acernese} F, {Ackley} K, {Adams} C,
  {Adams} T, {Addesso} P, {Adhikari} R~X, {Adya} V~B and et~al 2017 {\em
  Physical Review Letters\/} {\bf 119} 161101 (\textit{Preprint}
  \eprint{1710.05832})

\bibitem{Lonardoni:2015}
Lonardoni D, Lovato A, Gandolfi S and Pederiva F 2015 {\em Phys. Rev. Lett.\/}
  {\bf 114}(9) 092301
  \urlprefix\url{http://link.aps.org/doi/10.1103/PhysRevLett.114.092301}

\bibitem{Alford:2008}
Alford M~G, Schmitt A, Rajagopal K and Sch\"afer T 2008 {\em Rev. Mod. Phys.\/}
  {\bf 80}(4) 1455--1515
  \urlprefix\url{https://link.aps.org/doi/10.1103/RevModPhys.80.1455}

\bibitem{DAi16}
D'A\'{i} A, Evans P~A, Burrows D~N, Kuin N~P~M, Kann D~A, Campana S, Maselli A,
  Romano P, Cusumano G, La~Parola V, Barthelmy S~D, Beardmore A~P, Cenko S~B,
  De~Pasquale M, Gehrels N, Greiner J, Kennea J~A, Klose S, Melandri A, Nousek
  J~A, Osborne J~P, Palmer D~M, Sbarufatti B, Schady P, Siegel M~H, Tagliaferri
  G, Yates R and Zane S 2016 {\em Mon. Not. R. Astron. Soc.\/} {\bf 463}
  2394--2404 \urlprefix\url{https://doi.org/10.1093/mnras/stw2023}

\bibitem{Hessels06}
Hessels J~W~T, Ransom S~M, Stairs I~H, Freire P~C~C, Kaspi V~M and Camilo F
  2006 {\em Science\/} {\bf 311} 1901--1904
  \urlprefix\url{https://doi.org/10.1126/science.1123430}

\bibitem{Woods07}
Woods P~M, Kouveliotou C, Finger M~H, {G\"{o}\u{g}\'{o}\'{s}} E, Wilson C~A,
  Patel S~K, Hurley K and Swank J~H 2007 {\em Astrophys. J.\/} {\bf 654} 470
  \urlprefix\url{https://doi.org/10.1086/507459}

\bibitem{Karako-Argaman15}
Karako-Argaman C, Kaspi V~M, Lynch R~S, Hessels J~W~T, Kondratiev V~I,
  McLaughlin M~A, Ransom S~M, Archibald A~M, Boyles J, Jenet F~A, Kaplan D~L,
  Levin L, Lorimer D~R, Madsen E~C, Roberts M~S~E, Siemens X, Stairs I~H,
  Stovall K, Swiggum J~K and van Leeuwen J 2015 {\em Astrophys. J.\/} {\bf 809}
  67 (\textit{Preprint} \eprint{1503.05170})
  \urlprefix\url{https://doi.org/10.1088/0004-637X/809/1/67}

\bibitem{Steiner16ns}
Steiner A~W, Lattimer J~M and Brown E~F 2016 {\em Eur. Phys. J. A\/} {\bf 52}
  18 (\textit{Preprint} \eprint{1510.07515})
  \urlprefix\url{http://doi.org/10.1140/epja/i2016-16018-1}

\bibitem{Steiner15un}
Steiner A~W, Gandolfi S, Fattoyev F~J and Newton W~G 2015 {\em Phys. Rev. C\/}
  {\bf 91} 015804 (\textit{Preprint} \eprint{1403.7546})
  \urlprefix\url{https://doi.org/10.1103/PhysRevC.91.015804}

\bibitem{Steiner05ia}
Steiner A~W, Prakash M, Lattimer J~M and Ellis P~J 2005 {\em Phys. Rep.\/} {\bf
  411} 325 (\textit{Preprint} \eprint{nucl-th/0410066})
  \urlprefix\url{http://doi.org/10.1016/j.physrep.2005.02.004}

\bibitem{Akmal:1998}
Akmal A, Pandharipande V~R and Ravenhall D~G 1998 {\em Phys. Rev. C\/} {\bf
  58}(3) 1804--1828

\bibitem{Hebeler:2010}
Hebeler K and Schwenk A 2010 {\em Phys. Rev. C\/} {\bf 82} 014314

\bibitem{Hagen:2014}
Hagen G, Papenbrock T, Ekstr\"om A, Wendt K~A, Baardsen G, Gandolfi S,
  Hjorth-Jensen M and Horowitz C~J 2014 {\em Phys. Rev. C\/} {\bf 89}(1) 014319
  \urlprefix\url{https://link.aps.org/doi/10.1103/PhysRevC.89.014319}

\bibitem{Sarsa:2003}
Sarsa A, Fantoni S, Schmidt K~E and Pederiva F 2003 {\em Phys. Rev. C\/} {\bf
  68} 024308

\bibitem{Carlson:2003}
Carlson J, Morales J, Pandharipande V~R and Ravenhall D~G 2003 {\em Phys. Rev.
  C\/} {\bf 68}(2) 025802
  \urlprefix\url{https://link.aps.org/doi/10.1103/PhysRevC.68.025802}

\bibitem{Schmidt:1999}
Schmidt K~E and Fantoni S 1999 {\em Phys. Lett. B\/} {\bf 446} 99--103

\bibitem{Gandolfi:2015}
Gandolfi S, Gezerlis A and Carlson J 2015 {\em Annu. Rev. Nucl. Part. Sci.\/}
  {\bf 65} 303--328
  \urlprefix\url{http://dx.doi.org/10.1146/annurev-nucl-102014-021957}

\bibitem{Carlson:2015}
Carlson J, Gandolfi S, Pederiva F, Pieper S~C, Schiavilla R, Schmidt K~E and
  Wiringa R~B 2015 {\em Rev. Mod. Phys.\/} {\bf 87}(3) 1067--1118
  \urlprefix\url{http://link.aps.org/doi/10.1103/RevModPhys.87.1067}

\bibitem{Pudliner:1997}
Pudliner B~S, Pandharipande V~R, Carlson J, Pieper S~C and Wiringa R~B 1997
  {\em Phys. Rev. C\/} {\bf 56} 1720--1750 ISSN 0556-2813
  \urlprefix\url{http://link.aps.org/doi/10.1103/PhysRevC.56.1720}

\bibitem{Gandolfi:2009}
Gandolfi S, Illarionov A~Y, Schmidt K~E, Pederiva F and Fantoni S 2009 {\em
  Phys. Rev. C\/} {\bf {79}}(5) 054005
  \urlprefix\url{http://link.aps.org/doi/10.1103/PhysRevC.79.054005}

\bibitem{Lonardoni:2018prl}
Lonardoni D, Carlson J, Gandolfi S, Lynn J~E, Schmidt K~E, Schwenk A and Wang
  X~B 2018 {\em Phys. Rev. Lett.\/} {\bf 120}(12) 122502

\bibitem{Lonardoni:2018prc}
Lonardoni D, Gandolfi S, Lynn J~E, Petrie C, Carlson J, Schmidt K~E and Schwenk
  A 2018 {\em Phys. Rev. C\/} {\bf 97}(4) 044318

\bibitem{Lonardoni:2018nofk}
Lonardoni D, Gandolfi S, Wang X~B and Carlson J 2018 {\em Phys. Rev. C\/} {\bf
  98}(1) 014322

\bibitem{Wiringa:2002}
Wiringa R and Pieper S 2002 {\em Phys. Rev. Lett.\/} {\bf 89} 18--21 ISSN
  0031-9007
  \urlprefix\url{http://link.aps.org/doi/10.1103/PhysRevLett.89.182501}

\bibitem{Wiringa:1995}
Wiringa R~B, Stoks V~G~J and Schiavilla R 1995 {\em Phys. Rev. C\/} {\bf 51}(1)
  38--51 \urlprefix\url{http://link.aps.org/doi/10.1103/PhysRevC.51.38}

\bibitem{Pudliner:1995}
Pudliner B~S, Pandharipande V~R, Carlson J and Wiringa R~B 1995 {\em Phys. Rev.
  Lett.\/} {\bf 74} 4396--4399 ISSN 0031-9007
  \urlprefix\url{http://link.aps.org/doi/10.1103/PhysRevLett.74.4396}

\bibitem{Gandolfi:2012}
Gandolfi S, Carlson J and Reddy S 2012 {\em Phys. Rev. C\/} {\bf 85} 032801
  ISSN 0556-2813
  \urlprefix\url{http://link.aps.org/doi/10.1103/PhysRevC.85.032801}

\bibitem{Epelbaum:2009}
Epelbaum E, Hammer H~W and Mei\ss{}ner U~G 2009 {\em Rev. Mod. Phys.\/} {\bf
  81}(4) 1773--1825
  \urlprefix\url{http://link.aps.org/doi/10.1103/RevModPhys.81.1773}

\bibitem{Epelbaum:2015}
Epelbaum E, Krebs H and Mei\ss{}ner U~G 2015 {\em Eur. Phys. J. A\/} {\bf 51}
  53 \urlprefix\url{http://dx.doi.org/10.1140/epja/i2015-15053-8}

\bibitem{Melendez:2017phj}
Melendez J~A, Wesolowski S and Furnstahl R~J 2017 {\em Phys. Rev.\/} {\bf C96}
  024003 (\textit{Preprint} \eprint{1704.03308})

\bibitem{Gezerlis:2013}
Gezerlis A, Tews I, Epelbaum E, Gandolfi S, Hebeler K, Nogga A and Schwenk A
  2013 {\em Phys. Rev. Lett.\/} {\bf 111}(3) 032501
  \urlprefix\url{http://link.aps.org/doi/10.1103/PhysRevLett.111.032501}

\bibitem{Gezerlis:2014}
Gezerlis A, Tews I, Epelbaum E, Freunek M, Gandolfi S, Hebeler K, Nogga A and
  Schwenk A 2014 {\em Phys. Rev. C\/} {\bf 90} 054323

\bibitem{Lynn:2016}
Lynn J~E, Tews I, Carlson J, Gandolfi S, Gezerlis A, Schmidt K~E and Schwenk A
  2016 {\em Phys. Rev. Lett.\/} {\bf 116} 062501

\bibitem{Lonardoni:2018}
Lonardoni D, Carlson J, Gandolfi S, Lynn J~E, Schmidt K~E, Schwenk A and Wang
  X~B 2018 {\em Phys. Rev. Lett.\/} {\bf 120}(12) 122502
  \urlprefix\url{https://link.aps.org/doi/10.1103/PhysRevLett.120.122502}

\bibitem{Tews:2018}
{Tews} I, {Carlson} J, {Gandolfi} S and {Reddy} S 2018 {\em \apj\/} {\bf 860}
  149 (\textit{Preprint} \eprint{1801.01923})

\bibitem{Lattimer:1991nc}
Lattimer J~M and Swesty F~D 1991 {\em Nucl. Phys.\/} {\bf A535} 331--376

\bibitem{Hempel}
Hempel M private communication.

\bibitem{Shen}
Shen G private communication.

\bibitem{Typel}
Typel S private communication.

\bibitem{Steiner12cn}
Steiner A~W and Gandolfi S 2012 {\em Phys. Rev. Lett.\/} {\bf 108} 081102
  (\textit{Preprint} \eprint{1110.4142})
  \urlprefix\url{http://doi.org/10.1103/PhysRevLett.108.081102}

\bibitem{BPS:1971}
{Baym} G, {Pethick} C and {Sutherland} P 1971 {\em Astrophys. J.\/} {\bf 170}
  299--+

\bibitem{NegeleVautherin:1973}
{Negele} J~W and {Vautherin} D 1973 {\em Nucl. Phys. A\/} {\bf 207} 298--320

\bibitem{Read:2008iy}
Read J~S, Lackey B~D, Owen B~J and Friedman J~L 2009 {\em Phys. Rev.\/} {\bf
  D79} 124032 (\textit{Preprint} \eprint{0812.2163})

\bibitem{Hebeler:2013nza}
Hebeler K, Lattimer J~M, Pethick C~J and Schwenk A 2013 {\em Astrophys. J.\/}
  {\bf 773} 11 (\textit{Preprint} \eprint{1303.4662})

\bibitem{Steiner13tn}
Steiner A~W, Lattimer J~M and Brown E~F 2013 {\em Astrophys. J. Lett.\/} {\bf
  765} 5 (\textit{Preprint} \eprint{1205.6871})

\bibitem{Tews:2018kmu}
Tews I, Carlson J, Gandolfi S and Reddy S 2018 {\em Astrophys. J.\/} {\bf 860}
  149 (\textit{Preprint} \eprint{1801.01923})

\bibitem{Tews:2018chv}
Tews I, Margueron J and Reddy S 2018 {\em Phys. Rev.\/} {\bf C98} 045804
  (\textit{Preprint} \eprint{1804.02783})

\bibitem{Alford:2013aca}
Alford M~G, Han S and Prakash M 2013 {\em Phys. Rev.\/} {\bf D88} 083013
  (\textit{Preprint} \eprint{1302.4732})

\bibitem{Bedaque:2014sqa}
Bedaque P and Steiner A~W 2015 {\em Phys. Rev. Lett.\/} {\bf 114} 031103
  (\textit{Preprint} \eprint{1408.5116})

\bibitem{Tews:2016ofv}
Tews I 2017 {\em Phys. Rev.\/} {\bf C95} 015803 (\textit{Preprint}
  \eprint{1607.06998})

\bibitem{Tews:2019cap}
Tews I, Margueron J and Reddy S 2019  (\textit{Preprint} \eprint{1901.09874})

\bibitem{burrows:87b}
{Burrows} A and {Lattimer} J~M 1987 {\em \apjl\/} {\bf 318} L63

\bibitem{bruenn:87}
{Bruenn} S~W 1987 {\em \prl\/} {\bf 59} 938

\bibitem{hirata:87}
{Hirata} K, {Kajita} T, {Koshiba} M, {Nakahata} M and {Oyama} Y 1987 {\em
  \prl\/} {\bf 58} 1490

\bibitem{arnett:89}
{Arnett} W~D, {Bahcall} J~N, {Kirshner} R~P and {Woosley} S~E 1989 {\em
  \araa\/} {\bf 27} 629--700

\bibitem{Langanke03}
Langanke K, Martinez-Pinedo G, Sampaio J~M, Dean D~J, Hix W~R, Messer O~E~B,
  Mezzacappa A, Liebendoerfer M, Janka H~T and Rampp M 2003 {\em Phys. Rev.
  Lett.\/} {\bf 90} 241102 (\textit{Preprint} \eprint{astro-ph/0302459})
  \urlprefix\url{https://doi.org/10.1103/PhysRevLett.90.241102}

\bibitem{pagliaroli:09b}
{Pagliaroli} G, {Vissani} F, {Costantini} M~L and {Ianni} A 2009 {\em
  Astroparticle Physics\/} {\bf 31} 163--176 (\textit{Preprint}
  \eprint{0810.0466})

\bibitem{ligo:17nsns_dyn_ej}
{Abbott} B~P, {Abbott} R, {Abbott} T~D, {Acernese} F, {Ackley} K, {Adams} C,
  {Adams} T, {Addesso} P, {Adhikari} R~X, {Adya} V~B and et~al 2017 {\em
  \apjl\/} {\bf 850} L39 (\textit{Preprint} \eprint{1710.05836})

\bibitem{lattimer:76}
{Lattimer} J~M and {Schramm} D~N 1976 {\em \apj\/} {\bf 210} 549--567

\bibitem{mochkovitch:93}
{Mochkovitch} R, {Hernanz} M, {Isern} J and {Martin} X 1993 {\em \nat\/} {\bf
  361} 236

\bibitem{janka:99}
{Janka} H~T, {Eberl} T, {Ruffert} M and {Fryer} C~L 1999 {\em \apjl\/} {\bf
  527} L39--L42 (\textit{Preprint} \eprint{astro-ph/9908290})

\bibitem{freiburghaus:99}
{Freiburghaus} C, {Rosswog} S and {Thielemann} F~K 1999 {\em \apjl\/} {\bf 525}
  L121

\bibitem{wlee:07}
{Lee} W~H and {Ramirez-Ruiz} E 2007 {\em New Journal of Physics\/} {\bf 9} 17
  (\textit{Preprint} \eprint{astro-ph/0701874})

\bibitem{nakar:07a}
{Nakar} E 2007 {\em \physrep\/} {\bf 442} 166 (\textit{Preprint}
  \eprint{astro-ph/0701748})

\bibitem{gehrels:09}
{Gehrels} N, {Ramirez-Ruiz} E and {Fox} D~B 2009 {\em \araa\/} {\bf 47} 567

\bibitem{fong:13}
{Fong} W and {Berger} E 2013 {\em \apj\/} {\bf 776} 18 (\textit{Preprint}
  \eprint{1307.0819})

\bibitem{rosswog:15b}
{Rosswog} S 2015 {\em International Journal of Modern Physics D\/} {\bf 24}
  1530012-52 (\textit{Preprint} \eprint{1501.02081})

\bibitem{ligo:17nsnsmm}
{Abbott} B~P, {Abbott} R, {Abbott} T~D, {Acernese} F, {Ackley} K, {Adams} C,
  {Adams} T, {Addesso} P, {Adhikari} R~X, {Adya} V~B and et~al 2017 {\em
  \apjl\/} {\bf 848} L12 (\textit{Preprint} \eprint{1710.05833})

\bibitem{goldstein:17}
{Goldstein} A, {Veres} P, {Burns} E, {Briggs} M~S, {Hamburg} R, {Kocevski} D,
  {Wilson-Hodge} C~A, {Preece} R~D, {Poolakkil} S, {Roberts} O~J, {Hui} C~M,
  {Connaughton} V, {Racusin} J, {von Kienlin} A, {Dal Canton} T, {Christensen}
  N, {Littenberg} T, {Siellez} K, {Blackburn} L, {Broida} J, {Bissaldi} E,
  {Cleveland} W~H, {Gibby} M~H, {Giles} M~M, {Kippen} R~M, {McBreen} S,
  {McEnery} J, {Meegan} C~A, {Paciesas} W~S and {Stanbro} M 2017 {\em \apjl\/}
  {\bf 848} L14 (\textit{Preprint} \eprint{1710.05446})

\bibitem{savchenko:17}
{Savchenko} V, {Ferrigno} C, {Kuulkers} E, {Bazzano} A, {Bozzo} E, {Brandt} S,
  {Chenevez} J, {Courvoisier} T~J~L, {Diehl} R, {Domingo} A, {Hanlon} L,
  {Jourdain} E, {von Kienlin} A, {Laurent} P, {Lebrun} F, {Lutovinov} A,
  {Martin-Carrillo} A, {Mereghetti} S, {Natalucci} L, {Rodi} J, {Roques} J~P,
  {Sunyaev} R and {Ubertini} P 2017 {\em \apjl\/} {\bf 848} L15
  (\textit{Preprint} \eprint{1710.05449})

\bibitem{cowperthwaite:17}
{Cowperthwaite} P~S, {Berger} E, {Villar} V~A, {Metzger} B~D, {Nicholl} M,
  {Chornock} R, {Blanchard} P~K, {Fong} W, {Margutti} R, {Soares-Santos} M,
  {Alexander} K~D, {Allam} S, {Annis} J, {Brout} D, {Brown} D~A, {Butler} R~E,
  {Chen} H~Y, {Diehl} H~T, {Doctor} Z, {Drout} M~R, {Eftekhari} T, {Farr} B,
  {Finley} D~A, {Foley} R~J, {Frieman} J~A, {Fryer} C~L,
  {Garc{\'{\i}}a-Bellido} J, {Gill} M~S~S, {Guillochon} J, {Herner} K, {Holz}
  D~E, {Kasen} D, {Kessler} R, {Marriner} J, {Matheson} T, {Neilsen} Jr E~H,
  {Quataert} E, {Palmese} A, {Rest} A, {Sako} M, {Scolnic} D~M, {Smith} N,
  {Tucker} D~L, {Williams} P~K~G, {Balbinot} E, {Carlin} J~L, {Cook} E~R,
  {Durret} F, {Li} T~S, {Lopes} P~A~A, {Louren{\c c}o} A~C~C, {Marshall} J~L,
  {Medina} G~E, {Muir} J, {Mu{\~n}oz} R~R, {Sauseda} M, {Schlegel} D~J, {Secco}
  L~F, {Vivas} A~K, {Wester} W, {Zenteno} A, {Zhang} Y, {Abbott} T~M~C,
  {Banerji} M, {Bechtol} K, {Benoit-L{\'e}vy} A, {Bertin} E, {Buckley-Geer} E,
  {Burke} D~L, {Capozzi} D, {Carnero Rosell} A, {Carrasco Kind} M, {Castander}
  F~J, {Crocce} M, {Cunha} C~E, {D'Andrea} C~B, {da Costa} L~N, {Davis} C,
  {DePoy} D~L, {Desai} S, {Dietrich} J~P, {Drlica-Wagner} A, {Eifler} T~F,
  {Evrard} A~E, {Fernandez} E, {Flaugher} B, {Fosalba} P, {Gaztanaga} E,
  {Gerdes} D~W, {Giannantonio} T, {Goldstein} D~A, {Gruen} D, {Gruendl} R~A,
  {Gutierrez} G, {Honscheid} K, {Jain} B, {James} D~J, {Jeltema} T, {Johnson}
  M~W~G, {Johnson} M~D, {Kent} S, {Krause} E, {Kron} R, {Kuehn} K, {Nuropatkin}
  N, {Lahav} O, {Lima} M, {Lin} H, {Maia} M~A~G, {March} M, {Martini} P,
  {McMahon} R~G, {Menanteau} F, {Miller} C~J, {Miquel} R, {Mohr} J~J, {Neilsen}
  E, {Nichol} R~C, {Ogando} R~L~C, {Plazas} A~A, {Roe} N, {Romer} A~K,
  {Roodman} A, {Rykoff} E~S, {Sanchez} E, {Scarpine} V, {Schindler} R,
  {Schubnell} M, {Sevilla-Noarbe} I, {Smith} M, {Smith} R~C, {Sobreira} F,
  {Suchyta} E, {Swanson} M~E~C, {Tarle} G, {Thomas} D, {Thomas} R~C, {Troxel}
  M~A, {Vikram} V, {Walker} A~R, {Wechsler} R~H, {Weller} J, {Yanny} B and
  {Zuntz} J 2017 {\em \apjl\/} {\bf 848} L17 (\textit{Preprint}
  \eprint{1710.05840})

\bibitem{tanvir:17}
{Tanvir} N~R, {Levan} A~J, {Gonz{\'a}lez-Fern{\'a}ndez} C, {Korobkin} O,
  {Mandel} I, {Rosswog} S, {Hjorth} J, {D'Avanzo} P, {Fruchter} A~S, {Fryer}
  C~L, {Kangas} T, {Milvang-Jensen} B, {Rosetti} S, {Steeghs} D, {Wollaeger}
  R~T, {Cano} Z, {Copperwheat} C~M, {Covino} S, {D'Elia} V, {de Ugarte Postigo}
  A, {Evans} P~A, {Even} W~P, {Fairhurst} S, {Figuera Jaimes} R, {Fontes} C~J,
  {Fujii} Y~I, {Fynbo} J~P~U, {Gompertz} B~P, {Greiner} J, {Hodosan} G, {Irwin}
  M~J, {Jakobsson} P, {J{\o}rgensen} U~G, {Kann} D~A, {Lyman} J~D, {Malesani}
  D, {McMahon} R~G, {Melandri} A, {O'Brien} P~T, {Osborne} J~P, {Palazzi} E,
  {Perley} D~A, {Pian} E, {Piranomonte} S, {Rabus} M, {Rol} E, {Rowlinson} A,
  {Schulze} S, {Sutton} P, {Th{\"o}ne} C~C, {Ulaczyk} K, {Watson} D, {Wiersema}
  K and {Wijers} R~A~M~J 2017 {\em \apjl\/} {\bf 848} L27 (\textit{Preprint}
  \eprint{1710.05455})

\bibitem{tanaka:17}
{Tanaka} M, {Utsumi} Y, {Mazzali} P~A, {Tominaga} N, {Yoshida} M, {Sekiguchi}
  Y, {Morokuma} T, {Motohara} K, {Ohta} K, {Kawabata} K~S, {Abe} F, {Aoki} K,
  {Asakura} Y, {Baar} S, {Barway} S, {Bond} I~A, {Doi} M, {Fujiyoshi} T,
  {Furusawa} H, {Honda} S, {Itoh} Y, {Kawabata} M, {Kawai} N, {Kim} J~H, {Lee}
  C~H, {Miyazaki} S, {Morihana} K, {Nagashima} H, {Nagayama} T, {Nakaoka} T,
  {Nakata} F, {Ohsawa} R, {Ohshima} T, {Okita} H, {Saito} T, {Sumi} T,
  {Tajitsu} A, {Takahashi} J, {Takayama} M, {Tamura} Y, {Tanaka} I, {Terai} T,
  {Tristram} P~J, {Yasuda} N and {Zenko} T 2017 {\em \pasj\/} {\bf 69} 102
  (\textit{Preprint} \eprint{1710.05850})

\bibitem{coulter:17}
{Coulter} D~A, {Foley} R~J, {Kilpatrick} C~D, {Drout} M~R, {Piro} A~L,
  {Shappee} B~J, {Siebert} M~R, {Simon} J~D, {Ulloa} N, {Kasen} D, {Madore}
  B~F, {Murguia-Berthier} A, {Pan} Y~C, {Prochaska} J~X, {Ramirez-Ruiz} E,
  {Rest} A and {Rojas-Bravo} C 2017 {\em Science\/} {\bf 358} 1556--1558
  (\textit{Preprint} \eprint{1710.05452})

\bibitem{evans:17}
{Evans} P~A, {Cenko} S~B, {Kennea} J~A, {Emery} S~W~K, {Kuin} N~P~M, {Korobkin}
  O, {Wollaeger} R~T, {Fryer} C~L, {Madsen} K~K, {Harrison} F~A, {Xu} Y,
  {Nakar} E, {Hotokezaka} K, {Lien} A, {Campana} S, {Oates} S~R, {Troja} E,
  {Breeveld} A~A, {Marshall} F~E, {Barthelmy} S~D, {Beardmore} A~P, {Burrows}
  D~N, {Cusumano} G, {D'A{\`i}} A, {D'Avanzo} P, {D'Elia} V, {de Pasquale} M,
  {Even} W~P, {Fontes} C~J, {Forster} K, {Garcia} J, {Giommi} P, {Grefenstette}
  B, {Gronwall} C, {Hartmann} D~H, {Heida} M, {Hungerford} A~L, {Kasliwal} M~M,
  {Krimm} H~A, {Levan} A~J, {Malesani} D, {Melandri} A, {Miyasaka} H, {Nousek}
  J~A, {O'Brien} P~T, {Osborne} J~P, {Pagani} C, {Page} K~L, {Palmer} D~M,
  {Perri} M, {Pike} S, {Racusin} J~L, {Rosswog} S, {Siegel} M~H, {Sakamoto} T,
  {Sbarufatti} B, {Tagliaferri} G, {Tanvir} N~R and {Tohuvavohu} A 2017 {\em
  Science\/} {\bf 358} 1565--1570 (\textit{Preprint} \eprint{1710.05437})

\bibitem{li:98}
{Li} L~X and {Paczy{\'n}ski} B 1998 {\em \apjl\/} {\bf 507} L59
  (\textit{Preprint} \eprint{astro-ph/9807272})

\bibitem{lippuner:15}
{Lippuner} J and {Roberts} L~F 2015 {\em \apj\/} {\bf 815} 82
  (\textit{Preprint} \eprint{1508.03133})

\bibitem{fernandez:16}
{Fern{\'a}ndez} R and {Metzger} B~D 2016 {\em Annual Review of Nuclear and
  Particle Science\/} {\bf 66} annurev-nucl-102115-044819 (\textit{Preprint}
  \eprint{1512.05435})

\bibitem{metzger:17a}
{Metzger} B~D 2017 {\em Living Reviews in Relativity\/} {\bf 20} 3
  (\textit{Preprint} \eprint{1610.09381})

\bibitem{kasen:17}
{Kasen} D, {Metzger} B, {Barnes} J, {Quataert} E and {Ramirez-Ruiz} E 2017 {\em
  \nat\/} {\bf 551} 80--84 (\textit{Preprint} \eprint{1710.05463})

\bibitem{arnould:07}
{Arnould} M, {Goriely} S and {Takahashi} K 2007 {\em \physrep\/} {\bf 450} 97
  (\textit{Preprint} \eprint{0705.4512})

\bibitem{lippuner:17a}
{Lippuner} J, {Fern{\'a}ndez} R, {Roberts} L~F, {Foucart} F, {Kasen} D,
  {Metzger} B~D and {Ott} C~D 2017 {\em \mnras\/} {\bf 472} 904--918
  (\textit{Preprint} \eprint{1703.06216})

\bibitem{ligo:18_prospects}
{Abbott} B~P, {Abbott} R, {Abbott} T~D, {Abernathy} M~R, {Acernese} F, {Ackley}
  K, {Adams} C, {Adams} T, {Addesso} P, {Adhikari} R~X and et~al 2018 {\em
  Living Reviews in Relativity\/} {\bf 21} 3 (\textit{Preprint}
  \eprint{1304.0670})

\bibitem{scolnic:18}
{Scolnic} D, {Kessler} R, {Brout} D, {Cowperthwaite} P~S, {Soares-Santos} M,
  {Annis} J, {Herner} K, {Chen} H~Y, {Sako} M, {Doctor} Z, {Butler} R~E,
  {Palmese} A, {Diehl} H~T, {Frieman} J, {Holz} D~E, {Berger} E, {Chornock} R,
  {Villar} V~A, {Nicholl} M, {Biswas} R, {Hounsell} R, {Foley} R~J, {Metzger}
  J, {Rest} A, {Garc{\'\i}a-Bellido} J, {M{\"o}ller} A, {Nugent} P, {Abbott}
  T~M~C, {Abdalla} F~B, {Allam} S, {Bechtol} K, {Benoit-L{\'e}vy} A, {Bertin}
  E, {Brooks} D, {Buckley-Geer} E, {Carnero Rosell} A, {Carrasco Kind} M,
  {Carretero} J, {Castander} F~J, {Cunha} C~E, {D'Andrea} C~B, {da Costa} L~N,
  {Davis} C, {Doel} P, {Drlica-Wagner} A, {Eifler} T~F, {Flaugher} B, {Fosalba}
  P, {Gaztanaga} E, {Gerdes} D~W, {Gruen} D, {Gruendl} R~A, {Gschwend} J,
  {Gutierrez} G, {Hartley} W~G, {Honscheid} K, {James} D~J, {Johnson} M~W~G,
  {Johnson} M~D, {Krause} E, {Kuehn} K, {Kuhlmann} S, {Lahav} O, {Li} T~S,
  {Lima} M, {Maia} M~A~G, {March} M, {Marshall} J~L, {Menanteau} F, {Miquel} R,
  {Neilsen} E, {Plazas} A~A, {Sanchez} E, {Scarpine} V, {Schubnell} M,
  {Sevilla-Noarbe} I, {Smith} M, {Smith} R~C, {Sobreira} F, {Suchyta} E,
  {Swanson} M~E~C, {Tarle} G, {Thomas} R~C, {Tucker} D~L, {Walker} A~R and {DES
  Collaboration} 2018 {\em \apj\/} {\bf 852} L3 (\textit{Preprint}
  \eprint{1710.05845})

\bibitem{metzger:17b}
{Metzger} B~D 2017 {\em arXiv e-prints\/} (\textit{Preprint}
  \eprint{1710.05931})

\bibitem{rosswog:17}
{Rosswog} S, {Feindt} U, {Korobkin} O, {Wu} M~R, {Sollerman} J, {Goobar} A and
  {Martinez-Pinedo} G 2017 {\em Classical and Quantum Gravity\/} {\bf 34}
  104001 (\textit{Preprint} \eprint{1611.09822})

\bibitem{doctor:17}
{Doctor} Z, {Kessler} R, {Chen} H~Y, {Farr} B, {Finley} D~A, {Foley} R~J,
  {Goldstein} D~A, {Holz} D~E, {Kim} A~G, {Morganson} E, {Sako} M, {Scolnic} D,
  {Smith} M, {Soares-Santos} M, {Spinka} H, {Abbott} T~M~C, {Abdalla} F~B,
  {Allam} S, {Annis} J, {Bechtol} K, {Benoit-L{\'e}vy} A, {Bertin} E, {Brooks}
  D, {Buckley-Geer} E, {Burke} D~L, {Carnero Rosell} A, {Carrasco Kind} M,
  {Carretero} J, {Cunha} C~E, {D'Andrea} C~B, {da Costa} L~N, {DePoy} D~L,
  {Desai} S, {Diehl} H~T, {Drlica-Wagner} A, {Eifler} T~F, {Frieman} J,
  {Garc{\'{\i}}a-Bellido} J, {Gaztanaga} E, {Gerdes} D~W, {Gruendl} R~A,
  {Gschwend} J, {Gutierrez} G, {James} D~J, {Krause} E, {Kuehn} K, {Kuropatkin}
  N, {Lahav} O, {Li} T~S, {Lima} M, {Maia} M~A~G, {March} M, {Marshall} J~L,
  {Menanteau} F, {Miquel} R, {Neilsen} E, {Nichol} R~C, {Nord} B, {Plazas} A~A,
  {Romer} A~K, {Sanchez} E, {Scarpine} V, {Schubnell} M, {Sevilla-Noarbe} I,
  {Smith} R~C, {Sobreira} F, {Suchyta} E, {Swanson} M~E~C, {Tarle} G, {Walker}
  A~R, {Wester} W and {DES Collaboration} 2017 {\em \apj\/} {\bf 837} 57
  (\textit{Preprint} \eprint{1611.08052})

\bibitem{duez:10a}
{Duez} M~D, {Foucart} F, {Kidder} L~E, {Ott} C~D and {Teukolsky} S~A 2010 {\em
  \cqg\/} {\bf 27} 114106 (\textit{Preprint} \eprint{0912.3528})

\bibitem{hotokezaka:11}
{Hotokezaka} K, {Kyutoku} K, {Okawa} H, {Shibata} M and {Kiuchi} K 2011 {\em
  \prd\/} {\bf 83} 124008 (\textit{Preprint} \eprint{1105.4370})

\bibitem{deaton:13}
{Deaton} M~B, {Duez} M~D, {Foucart} F, {O'Connor} E, {Ott} C~D, {Kidder} L~E,
  {Muhlberger} C~D, {Scheel} M~A and {Szilagyi} B 2013 {\em \apj\/} {\bf 776}
  47 (\textit{Preprint} \eprint{1304.3384})

\bibitem{bauswein:14}
{Bauswein} A, {Stergioulas} N and {Janka} H~T 2014 {\em \prd\/} {\bf 90} 023002
  (\textit{Preprint} \eprint{1403.5301})

\bibitem{palenzuela:15}
{Palenzuela} C, {Liebling} S~L, {Neilsen} D, {Lehner} L, {Caballero} O~L,
  {O'Connor} E and {Anderson} M 2015 {\em \prd\/} {\bf 92} 044045
  (\textit{Preprint} \eprint{1505.01607})

\bibitem{kawaguchi:15}
{Kawaguchi} K, {Kyutoku} K, {Nakano} H, {Okawa} H, {Shibata} M and {Taniguchi}
  K 2015 {\em \prd\/} {\bf 92} 024014 (\textit{Preprint} \eprint{1506.05473})

\bibitem{radice:16b}
{Radice} D, {Galeazzi} F, {Lippuner} J, {Roberts} L~F, {Ott} C~D and {Rezzolla}
  L 2016 {\em \mnras\/} {\bf 460} 3255 (\textit{Preprint} \eprint{1601.02426})

\bibitem{foucart:17a}
{Foucart} F, {Desai} D, {Brege} W, {Duez} M~D, {Kasen} D, {Hemberger} D~A,
  {Kidder} L~E, {Pfeiffer} H~P and {Scheel} M~A 2017 {\em Classical and Quantum
  Gravity\/} {\bf 34} 044002 (\textit{Preprint} \eprint{1611.01159})

\bibitem{radice:18}
{Radice} D, {Perego} A, {Hotokezaka} K, {Fromm} S~A, {Bernuzzi} S and {Roberts}
  L~F 2018 {\em \apj\/} {\bf 869} 130 (\textit{Preprint} \eprint{1809.11161})

\bibitem{kyutoku:18}
{Kyutoku} K, {Kiuchi} K, {Sekiguchi} Y, {Shibata} M and {Taniguchi} K 2018 {\em
  \prd\/} {\bf 97} 023009 (\textit{Preprint} \eprint{1710.00827})

\bibitem{martin:15}
{Martin} D, {Perego} A, {Arcones} A, {Thielemann} F~K, {Korobkin} O and
  {Rosswog} S 2015 {\em \apj\/} {\bf 813} 2 (\textit{Preprint}
  \eprint{1506.05048})

\bibitem{wanajo:14}
{Wanajo} S, {Sekiguchi} Y, {Nishimura} N, {Kiuchi} K, {Kyutoku} K and {Shibata}
  M 2014 {\em \apjl\/} {\bf 789} L39 (\textit{Preprint} \eprint{1402.7317})

\bibitem{metzger:14}
{Metzger} B~D and {Fern{\'a}ndez} R 2014 {\em \mnras\/} {\bf 441} 3444
  (\textit{Preprint} \eprint{1402.4803})

\bibitem{sekiguchi:15}
{Sekiguchi} Y, {Kiuchi} K, {Kyutoku} K and {Shibata} M 2015 {\em \prd\/} {\bf
  91} 064059 (\textit{Preprint} \eprint{1502.06660})

\bibitem{foucart:15}
{Foucart} F, {O'Connor} E, {Roberts} L, {Duez} M~D, {Haas} R, {Kidder} L~E,
  {Ott} C~D, {Pfeiffer} H~P, {Scheel} M~A and {Szilagyi} B 2015 {\em \prd\/}
  {\bf 91} 124021 (\textit{Preprint} \eprint{1502.04146})

\bibitem{goriely:15}
{Goriely} S, {Bauswein} A, {Just} O, {Pllumbi} E and {Janka} H~T 2015 {\em
  \mnras\/} {\bf 452} 3894--3904 (\textit{Preprint} \eprint{1504.04377})

\bibitem{foucart:16a}
{Foucart} F, {O'Connor} E, {Roberts} L, {Kidder} L~E, {Pfeiffer} H~P and
  {Scheel} M~A 2016 {\em \prd\/} {\bf 94} 123016 (\textit{Preprint}
  \eprint{1607.07450})

\bibitem{roberts:17b}
{Roberts} L~F, {Lippuner} J, {Duez} M~D, {Faber} J~A, {Foucart} F, {Lombardi}
  Jr J~C, {Ning} S, {Ott} C~D and {Ponce} M 2017 {\em \mnras\/} {\bf 464} 3907
  (\textit{Preprint} \eprint{1601.07942})

\bibitem{siegel:17}
{Siegel} D~M and {Metzger} B~D 2017 {\em ArXiv e-prints\/} (\textit{Preprint}
  \eprint{1705.05473})

\bibitem{cote:18}
{C{\^o}t{\'e}} B, {Fryer} C~L, {Belczynski} K, {Korobkin} O,
  {Chru{\'s}li{\'n}ska} M, {Vassh} N, {Mumpower} M~R, {Lippuner} J, {Sprouse}
  T~M, {Surman} R and {Wollaeger} R 2018 {\em \apj\/} {\bf 855} 99
  (\textit{Preprint} \eprint{1710.05875})

\bibitem{Oertel17eo}
Oertel M, Hempel M, Kl\"ahn T and Typel S 2017 {\em Rev. Mod. Phys.\/} {\bf
  89}(1) 015007
  \urlprefix\url{https://link.aps.org/doi/10.1103/RevModPhys.89.015007}

\bibitem{Schneider17}
Schneider A~S, Roberts L~F and Ott C~D 2017 {\em Phys. Rev.\/} {\bf C96} 065802
  \urlprefix\url{https://doi.org/10.1103/PhysRevC.96.065802}

\bibitem{Du18hd}
Du X, Steiner A~W and Holt J~W 2019 {\em Phys. Rev. C\/} {\bf 99} 025803
  \urlprefix\url{https://doi.org/10.1103/PhysRevC.99.025803}

\bibitem{Horowitz06}
Horowitz C~J and Schwenk A 2006 {\em Phys. Lett. B\/} {\bf 638} 153
  \urlprefix\url{https://doi.org/10.1016/j.physletb.2006.05.055}

\bibitem{Kortelainen14}
Kortelainen M, McDonnell J, Nazarewicz W, Olsen E, Reinhard P~G, Sarich J,
  Schunck N, Wild S~M, Davesne D, Erler J and Pastore A 2014 {\em Phys. Rev.
  C\/} {\bf 89} 054314
  \urlprefix\url{https://doi.org/10.1103/PhysRevC.89.054314}

\bibitem{Wellenhofer+14}
Wellenhofer C, Holt J~W, Kaiser N and Weise W 2014 {\em Phys. Rev. C\/} {\bf
  89} 064009 \urlprefix\url{https://doi.org/10.1103/PhysRevC.89.064009}

\bibitem{Wellenhofer15to}
Wellenhofer C, Holt J~W and Kaiser N 2015 {\em Phys. Rev. C\/} {\bf 92} 015801
  \urlprefix\url{https://doi.org/10.1103/PhysRevC.92.015801}

\bibitem{Damour:2009vw}
Damour T and Nagar A 2009 {\em Phys. Rev.\/} {\bf D80} 084035
  (\textit{Preprint} \eprint{0906.0096})

\bibitem{AbbottPRL:2017}
Abbott B~P, Abbott R, Abbott T~D, Acernese F, Ackley K, Adams C, Adams T,
  Addesso P, Adhikari R~X, Adya V~B, Affeldt C, Afrough M, Agarwal B, Agathos
  M, Agatsuma K {\em et~al.\/} (LIGO Scientific Collaboration and Virgo
  Collaboration) 2017 {\em Phys. Rev. Lett.\/} {\bf 119}(16) 161101

\bibitem{De:2018uhw}
De S, Finstad D, Lattimer J~M, Brown D~A, Berger E and Biwer C~M 2018 {\em
  Phys. Rev. Lett.\/} {\bf 121} 091102 [Erratum: Phys. Rev.
  Lett.121,no.25,259902(2018)] (\textit{Preprint} \eprint{1804.08583})

\bibitem{Abbott:2018wiz}
Abbott B~P {\em et~al.\/} (LIGO Scientific, Virgo) 2019 {\em Phys. Rev.\/} {\bf
  X9} 011001 (\textit{Preprint} \eprint{1805.11579})

\bibitem{Ozel09rt}
\"{O}zel F and Psaltis D 2009 {\em Phys. Rev. D\/} {\bf 80} 103003
  \urlprefix\url{https://doi.org/10.1103/PhysRevD.80.103003}

\bibitem{Alford13gc}
Alford M~G, Han S and Prakash M 2013 {\em Phys. Rev.\/} {\bf D88} 083013
  \urlprefix\url{https://doi.org/10.1103/PhysRevD.88.083013}

\bibitem{Han18td}
Han S and Steiner A~W 2018  (\textit{Preprint} \eprint{1810.10967})

\bibitem{Riley18op}
Riley T~E, Raaijmakers G and Watts A~L 2018 {\em Mon. Not. Roy. Astron. Soc.\/}
  {\bf 478} 1093--1131 \urlprefix\url{https://doi.org/10.1093/mnras/sty1051}

\bibitem{Steiner18ta}
Steiner A~W 2018  (\textit{Preprint} \eprint{1802.05339})

\end{thebibliography}

\providecommand{\newblock}{}

\end{document}